# A Survey of State Management in Big Data Processing Systems

Quoc-Cuong To[1] Juan Soto[1,2] Volker Markl[1,2]

[1] German Research Center for Artificial Intelligence (DFKI)
Alt-Moabit 91c
10559 Berlin, Germany
quoc_cuong.to@dfki.de

juan.soto@tu-berlin.de

[2] Technische Universität Berlin
FG DIMA, Sekr. EN-7 Raum 728,
Einsteinufer 17
10587 Berlin, Germany
volker.markl@tu-berlin.de

## ABSTRACT

The concept of *state* and its applications vary widely across big data processing systems. This is evident in both the research literature and existing systems, such as Apache Flink, Apache Heron, Apache Samza, Apache Spark, and Apache Storm. Given the pivotal role that state management plays, particularly, for *iterative* batch and stream processing, in this survey, we present examples of state as an enabler, discuss the alternative approaches used to handle and implement state, capture the many facets of state management, and highlight new research directions. Our aim is to provide insight into disparate state management techniques, motivate others to pursue research in this area, and draw attention to open problems.

## CCS Concepts

**computing methodologies; database management systems; information systems; massively parallel and high-performance computer systems**

## Keywords

big data processing systems, state management, survey

## 1. Introduction

Big data systems process massive amounts of data efficiently, often with fast response times and are typically characterized by the 4V's [28, 124], i.e., volume, variety, velocity, and veracity. In addition, they are generally classified by their data processing approach, i.e., *batch*-oriented *vs.* *stream*-oriented. In batch-oriented systems, processing occurs on chunks of large data files, whereas in stream-oriented systems, processing happens on continuously arriving data.

One of the first proposals for parallel batch-oriented data processing (BDP) was MapReduce [24], which became popularized via Hadoop, an open source-framework, due to its features, including *flexibility*, *fault-tolerance*, *programming ease*, and *scalability*. Today, it is widely regarded as the pioneer for large-scale data analysis. However, despite its merits, MapReduce has several drawbacks, such as a low-level programming model and a lack of support for iterations, which severely affects both the ease of use and performance, as well as its inability to deal with data streams. Consequently, alternatives were proposed to overcome these limitations. Among them were the BDP approaches surveyed by Doulkeridis et al. [28]. Additionally, novel scalable stream processing solutions, such as Apache Flink [19, 134] (a Stratosphere fork [6]), Apache Heron [130, 135], and Apache Spark [137] arose to meet the needs of an ever-increasing number of real-time applications demanding both *low latency* and *high throughput*.

Big data processing systems encompass a wide range of concepts, such as data flow operators, distributed scale out, and fault-tolerance, all of which leverage, manage and/or manipulate state. Data analytics programs can be modeled as directed data flow graphs or trees (in the absence of iterations or shared results). From this perspective, the analysis results are the *roots*, operators are the *intermediate nodes*, and data are the *leaves*. Each operator node performs an operation that transforms inputs flowing through it into outputs. Data flows from the leaves through the operator nodes to the roots.

Operators come in two varieties. *Stateless operators* are purely functional and they produce output, solely based on their input. Examples of stateless operators include relational selection, relational projection without duplicate elimination, or merging two inputs. In contrast, *stateful operators* compute their output on a sequence of inputs and potentially use additional side information, maintained in an internal data structure called *state*. Roy and Haridi [123] define *state* to be "*a sequence of values in time that contain the intermediate results of a desired computation.*" This construct preserves the history of past operations and affects the processing logic in subsequent computations. Examples of *stateful operators* include sorting, relational joins, or aggregation. Note: We will introduce our own definition of *state* later in Section 2.

Large-scale BDP frameworks that employ a functional programming paradigm, such as MapReduce, forbid programmers from using state explicitly due to their focus on scale out through parallelism. In particular, iterative computation suffers from this conceptual limitation, as one cannot efficiently leverage state re-use among different executions of a step function (i.e., the function being repeatedly executed) during an iteration. Approaches that incorporate state in a functional model, include online MapReduce systems [23] and Twister [29], "[which] *can result in custom, fragile code and disappointing performance*," as stated by Logothetis et al. [67].

On the other hand, stream processing frameworks incorporate *state*, to discretize continuous data streams and apply computations on subsets. Researchers have proposed novel ways to represent, manage, and use state in scalable data stream processing. For example, *windowing* is the main abstraction used to discretize data streams, as reflected by Matteis and Mencagli [72]. Alternatively, Fernandez et al. [36] propose using data structures, such as *key-value pairs* to represent the various state types (e.g., processing state, buffer state, routing state). These are discussed later in subsection 2.2.

State management has received much attention in recent years. Systems researchers are arduously working on addressing several key questions, including "*How to efficiently handle state in varying scenarios?*" and "*How can state be used across applications?*" This survey examines leading research across the foremost publications that address varying *concepts* arising in *state management* and particular *applications that depend on the use of state*. Figure 1 structures *state management* into five concepts (i.e., operations, incremental maintenance, state sharing, load balancing and elasticity, and performance) and three applications of state (i.e., stateful computation, iterative processing, and fault tolerance), each according to key question they address. Each of these eight facets is addressed in the subsequent sections.

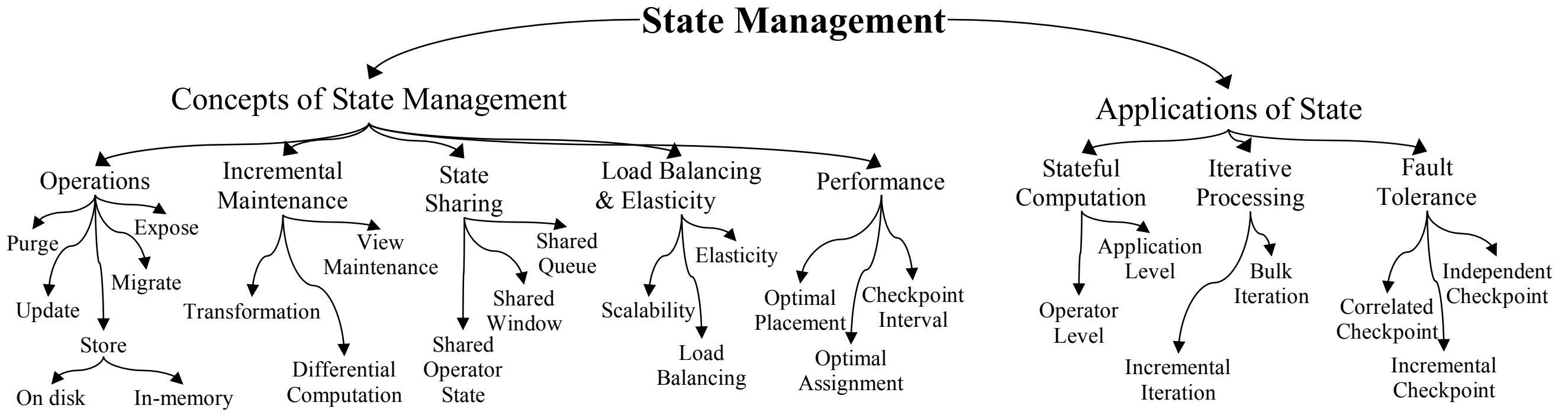

**Figure 1**. Diverse facets of state management.

The rest of this survey is structured as follows. In Section 2, the scope of the survey, the varying types of state, and related work are specified. In Section 3, we discuss the concepts of state management, including operations, incremental maintenance, state sharing, load balancing and elasticity as well as performance. In Section 4, we present an overview of how state is used as an enabler in important applications, i.e., stateful computation, iterative computation, and fault tolerance. In Section 5, we introduce integrative optimization, a cross-cutting topic spanning multiple concepts that is not explicitly represented in Figure 1. In Section 6, the implementation of state in today's leading big data processing frameworks, such as Flink [134], Heron [135], Samza [136], and Spark [137] as well as their limitations are examined. In Section 7, promising new research directions are underscored. And finally, in Section 8, closing remarks are offered.

# 2. Scope, Types of State, and Related Work

In this section, we specify the scope of the survey and introduce the varying types of state. In addition, we highlight related work that is out of scope.

## 2.1 Scope

In computer science, the *state* of a system arises in various domains, including programming languages, compilers, transfer protocols, formal specification, and data management. Given the broad nature of this topic, the scope of this survey is limited to state in big data management systems, in particular considering database and distributed systems centric research that largely focuses on states that may not necessarily fit into main memory and/or are distributed, partitioned, or replicated. With this in mind, we define *state* to be "*the intermediate value of a specific computation that will be used in subsequent operations during the processing of a data flow.*" We should note that this definition differs from its common use in traditional database systems, where *state* is a set of relational tables at a specific point in time. In some big data processing systems, large state sizes can be stored in either database systems or file systems, such as Cassandra, RocksDB, GFS, and HDFS.

Our focus is on the varying state management methods that have been published at top-tier venues in the big data domain over the past years. Our aim is to organize and synthesize the latest ideas in state management and layout some promising research directions in this domain. This survey is designed to enable readers to quickly grasp the state-of-the-art (SOTA) in state management, leverage and incorporate existing results into their own work, and encourage systems researchers to contribute novel ideas to advance the SOTA.

## 2.2 Types of State

State has various representations across big data processing systems. In this section, we describe the types of state, relevant in the survey from varying viewpoints. There is an *operator view*, where processing state, buffer state, and routing state belong. There is a *system view*, where there are computation state and configuration state. There is also an *application view*, where there are query state and program state. Lastly, there is a *programming view*, where there are variable state and window state. These are all depicted in **Table 1**. Next, we delve into each of these views.

| Views of State | Types of State |
|---|---|
| System View | Configuration State, Computation State |
| Application View | Query State, Program State |
| Programming View | Window State, Variable State |
| Operator View | Processing State, Routing State, Buffer State |

**Table 1**. Types of state classified by view.

**Operator View.** *Operator state* [36] is the most common type of state used in big data processing systems. It specifies the status of an operator and consists of several components, including *processing state*, input/output *buffer state* (a.k.a., input/output queue), and *routing state*. Using efficient data structures, processing state maintains an internal summary of the input (e.g., records) history. When necessary, systems translate processing state into an external serialized format (e.g., key-value pairs). Buffer state is realized by an operator's output buffer, which stores records that have not yet been processed (i.e., a limited number of output tuples from the past). In the papers we surveyed, upstream operators must cache these tuples, so that downstream operators can reprocess them upon failure. Using this caching mechanism, buffer state absorbs short-term variations of input rates and network bandwidth. After dynamic scale out, tuples must be delivered from an output buffer to an exact partitioned downstream operator. To do so, systems rely on routing state to direct a single tuple to a suitable partitioned downstream operator via key mappings.

**System View.** There are other less commonly used definitions of state. For example, in ChronoStream [111], the authors propose two types of state, i.e., *computation state* and *configuration state*. Computation state is "*a collection of application-level data structures that can be directly accessed and manipulated according* *to user-defined execution logic*." Configuration state is "*the set of container-level states that maintains the runtime-relevant parameters*."

**Application and Programming Views.** From the application viewpoint, there are *query state* and *program state*. In SEEP [36], query state consists of the state of each query operator. In GraphLab [69], program state is the compact representation of the program execution in a directed graph. From the programming viewpoint, there are *window state* and *variable state*. In S-Store [75], window state contains a finite, uninterrupted sequence of stream values. In CAPSULE [68], variable state is a data structure at the programming-language level for specific scenarios (e.g., checkpointing operator state for passive standby) in streaming applications. Similar definitions of state can be found in other applications and programming abstractions.

## 2.3 Related Work

Although there are numerous related works addressing the concepts of state management and the applications of state, there are several shortcomings. Some works reference *state*, yet their use of the term differs from our own. For example, in HTTP (the Hypertext Transfer Protocol) cookies (i.e., states) store historical information in web browsers, however, these do not play a role in data flow computations. Yet another example arises in automata theory, where formal specifications describe a finite state machine, where in this context state takes on a different meaning.

Existing Surveys on big data systems [28, 124] do not focus on state management per se, but rather discuss different aspects, such as *data storage*, *redundant processing*, or *join operations*. Fernandez et al. [37] give a brief overview on state, but do not discuss key aspects, such as load balancing and elasticity. To the best of our knowledge, our survey is the first to address the various facets of state management in big data processing systems.

# 3. Concepts of State Management

In this section, we discuss the five concepts of state management that we chose to focus on. That is, *operations* on state, *state sharing*, *incremental* state *maintenance*, *load balancing and elasticity*, and *performance* considerations. Note: Since some methods address multiple concepts, they are discussed once again from another perspective in subsequent subsections.

## 3.1 Operations

Handling state efficiently presents numerous technical challenges. For example, state can be *migrated among operators or nodes in a cluster* [26, 35] and *exposed to programmers for easier use* [36, 111], *maintained incrementally* [33] to improve performance, *shared among different processes* [15] to save storage, *stored remotely or locally*, using in-memory [83, 92] or disk spilling [63] techniques and balancing system load, potentially *even geographically distributed* [8]. There are many operations on state, including *store*, *update*, *purge*, *migrate*, and *expose*. In the following subsections, we discuss each of these operations and their impact on *state* in greater detail.

### *3.1.1 Storing State*

Storage solutions for *state* vary widely and generally *state size* determines, where state will be stored. For small sizes, researchers [92, 118] propose storing state *in-memory*, which can accelerate processing [118], but can also impact recovery efforts from machine failures. In this case, replicating the state to different machines will be needed, in order to recover from even transient machine failures. In contrast, for large sizes, researchers [57, 63, 83] have developed solutions, where state is kept in *persistent storage*. However, this incurs greater overhead. Nonetheless, deciding where to optimally store state is not always trivial. Next, we discuss three state handling solutions for large state sizes, i.e., *load shedding*, *state spilling*, and *state cleanup delay*.

Processing long-running queries (LRQ) over data streams (i.e., complex queries with huge operator states, such as multi-joins) can be memory-intensive. When system resources are scarce and processing demands cannot be met (e.g., due to high throughput and insufficient storage or compute capacity), varying handling methods can be employed. For example, *load shedding* [103] preserves just a subset of the state (e.g., as a sample, synopsis, or by lossy compression), which reduces workloads and increases performance, but at the expense of lowering accuracy. Workloads can be shed permanently or alternatively processed later when computing resources are again available [63].

For those cases where accuracy is paramount, load shedding is not a viable solution. Thus, an alternative approach, called state spilling, can be employed. This is true in particular for stateful relational operators (e.g., join variants, such as Hash-Merge Join [78], XJoin [106], and MJoin [107]), which temporarily flush states stored *in-memory* to disks when memory is at capacity. Yet another option is delaying state cleanup (i.e., processing states stored on disks) until resources are readily available. Each of these state handling solutions achieves both low-latency processing and the accuracy of results. Next, we present four approaches for storing and checkpointing state for fault-tolerance purposes.

The first approach due to Liu et al. [63], addresses the LRQ problem. Unlike existing solutions, which can only handle a single state-intensive operator in a data flow, such as a join operator, their strategies can handle multiple state-intensive operators. These multiple state-intensive operators arise in particular in data integration and data warehouse scenarios, where memory intensive queries abound. Their state spilling strategies selectively flush *operator states* to disks, to cope with complex queries. By appropriately spilling parts of operator state to disk at runtime, they avoid memory overflows and increase query throughput.

In addition, they observe that by exploiting operator interdependencies they can achieve higher performance over existing strategies. Further, they highlight two classes of data spilling strategies, namely, *operator-level* and *partition-level*. The operator-level strategy employs a bottom-up approach and regards all data in an operator state to be similarly important. In contrast, each of the partition-level data spill strategies (i.e., *local output*, *global output*, and *global output with penalty*) takes input data characteristics into account. In all of these strategies, when memory is scant, the appropriate partition to be spilled will need to be selected, to maximize query throughput.

The second approach due to Kwon et al., called *SGuard* [57], stores state in a distributed and replicated file system (DFS), such as the Google File System (GFS) and Hadoop Distributed File System (HDFS), to save memory for critical stream processing operations. One of the benefits of these file systems is that they are optimized for reading and writing large data volumes in bulk. Since multiple nodes may write state simultaneously, resolving resource conflicts is a critical requirement, which is met in SGuard by incorporating a scheduler into the DFS. The coordination of many write requests, enables the scheduler to reduce both individual checkpoint times and generally provides good resource utilization.

Akin to rollback recovery methods [49], SGuard periodically checkpoints state and recovers failed nodes from their last checkpoints. Unlike previous approaches, however, SGuard checkpoints asynchronously: While the system is under execution, SGuard uses a new *Memory Management Middleware* to store the operator state. As a consequence, this asynchronous mechanism can prevent potential interrupts and reduce the overhead incurred by the checkpointing process.

The third approach due to Nicolae et al. [83], proposes an asynchronous checkpointing runtime approach, called *AI-Ckpt*, designed for adaptive incremental state storing. AI-Ckpt exploits trends in *current* and *past* access patterns and generates an optimal ordering scheme to flush memory pages to stable storage. In their research paper, the authors observe that there are memory writing patterns in iterative applications. Consequently, AI-Ckpt leverages these patterns and optimizes the system to flush modified pages with minimum overhead.

Their experiments show that flushing optimally can considerably improve performance, especially for iterative applications (e.g., graph algorithms, machine learning) that exhibit repetitive access patterns. However, this method only uses the access order to flush pages and omits temporal aspects. Thus, a promising research direction is the integration of the timestamps and access order, in order to further improve the page flushing process.

Lastly, the fourth approach due to Ananthanarayanan et al. [8], called *Photon*, is a distributed system that can store large states across geographically distant locations. It can join multiple unordered data streams to ensure high scalability, low latency, and exactly-once semantics. Without human involvement, Photon can automatically solve infrastructure breakdowns and server outages. The critical state stored in the *IdRegistry* and shared between workers consists of a set of event identifiers (i.e., identifiers assigned to events), joined over the last *N* days, where *N* is chosen such that it balances storage costs and drop events. To ensure services are always available, the IdRegistry is duplicated synchronously across multiple datacenters, which may be in different geographical regions.

### *3.1.2 Updating State*

In this subsection, we turn our attention to four concepts to update state. That is, *incremental state update*, *fine-grained update*, *consistent update*, and *update semantics*.

In the first approach, Logothetis et al. [67] handle continuous bulk processing (CBP), by strictly updating a fragment of the *state* to optimize system performance. Similarly, Fegaras [33] updates *state* incrementally, via a new stateful operator, called *Incr*. Every time the MRQL (pronounced miracle) Streaming system produces a small delta result based on a data subset ($\Delta S_i$) and involving a homomorphism, it merges the previous state value and the current delta result, then the system can incrementally produce a new state value, i.e., $state \leftarrow state \otimes h(\Delta S_i)$. Figure 2 illustrates this update with two streaming sources.

In the second approach, Fernandez et al. [37] consider *fine-grained updates* to examine how updates can affect throughput and latency. They compare the update granularity among several systems to determine which one can support fine-grained updates. To do this, they vary the window size, since it depends on the granularity of updates to the state. That is, the smaller window size leads to less batching and thus finer granularity. Their experiments show that Naiad [79] can achieve low latency when using small batch sizes (e.g., 1000 messages) and high throughput for large batch sizes (e.g., 20000 messages). This result is due to Naiad's capability to configure the batch size, which is independent of the window size. Stateful dataflow graphs (SDG) [37] handle all window sizes and achieve higher throughput than Naiad. The overhead of micro-batching is substantial in other deployments: Spark Streaming throughput is equivalent to that of a SDG, but its smallest window size is 250 ms. If this limit is surpassed, its throughput will collapse.

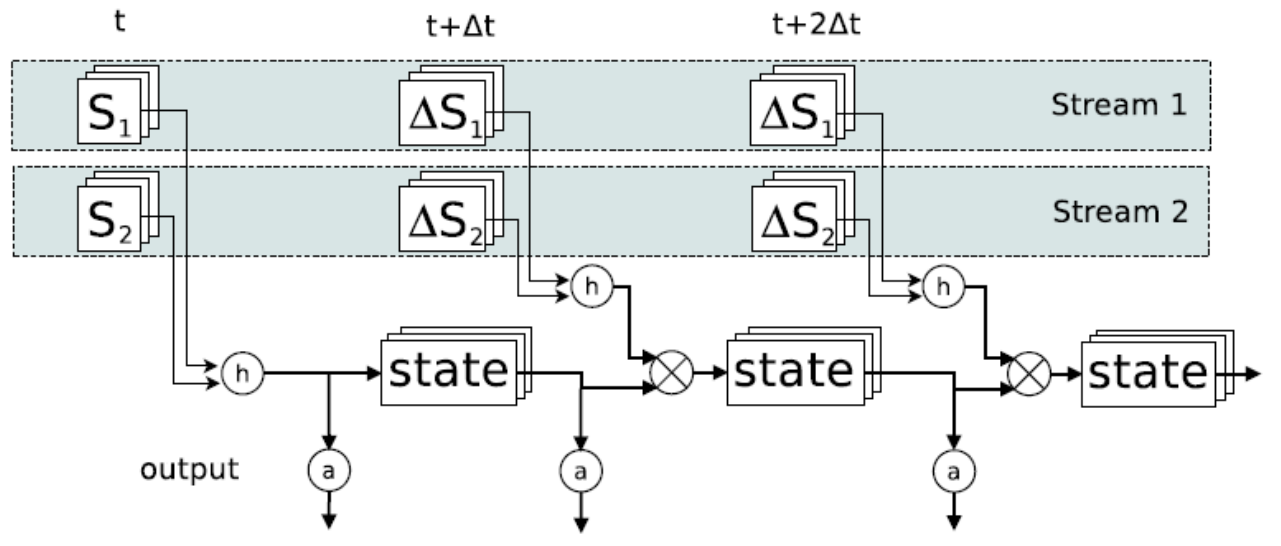

**Figure 2**. Incremental updates to *state* [33].

In the third approach, Low et al. [69] introduce the *GraphLab* framework for graph-parallel computation, to ensure data consistency when updating program *state*. GraphLab represents modifiable program state as a directed graph, called a *data graph*. This state includes user-defined mutable data and sparse computational dependencies. To alter the state, an update function transforms the graph into *scopes*, which are small overlapping contexts. To preserve data consistency, GraphLab presents three consistency models: *full, edge*, and *vertex* for update functions (UF).

These models enable the optimization of parallel execution and select the consistency level needed for correctness. The full consistency model achieves *serializability* by ensuring that the scopes of UF do not overlap and that the UF are executed concurrently. However, this consistency model limits potential parallelism and thus they propose two other consistency models to overcome this shortcoming. In the edge consistency model, each update function can read or write to its adjacent edges and vertex, but can only read adjacent vertices. Finally, all update functions can run in parallel in the vertex consistency model. As a result, these two consistency models improve parallelism.

Lastly, in the fourth approach, several big data processing frameworks [6, 102, 116] both explore and compare different *update semantics for state.* Basically, there are three types of semantic guarantees, namely, *at-least-once, at-most-once, and exactly-once*, to assess the correctness of state. Systems with *at-least-once* semantics fully process every tuple, but they cannot guarantee duplications in processing and thus addition of a tuple to the state. In *at-most-once* semantics, systems either do not process a tuple at all or execute an operation and add it to the state exactly once. Unlike *at-least-once* semantics, *at-most-once* semantics do not require the detection of duplicate tuples. Finally, systems with *exactly-once* semantics process tuples once and only once, thereby providing the strongest guarantee. In Section 6, we compare these semantic guarantees among popular big data frameworks.

### *3.1.3 Purging State*

When systems no longer need a specific piece of data for subsequent operations, state management can *purge* that data (e.g., a buffer state removing expired tuples). This subsection presents three efficient ways to purge state.

In the first approach, Ding et al. [25] propose several join algorithms that effectively purge state using punctuation on data attributes. They introduce a stream join operator, called *PJoin*, that deletes data, which is no longer useful. The use of punctuations marks the end of transmission values, thereby allowing stateful operators to remove state during runtime. Consequently, this frees memory for other operations and accelerates the probing process in join operations. Then, they equip PJoin with two strategies, i.e., eager and lazy purging. *Eager purge* immediately purges states whenever punctuations are observed, to minimize memory overhead and efficiently probe the state of the join operation. If punctuations arrive too frequently, then eager purge is not applicable since the probing cost is less than the cost of scanning the join state. Therefore, they propose *lazy (batch) purge*, which can only initiate purging when the number of newly generated punctuations from the last purge approaches a given threshold. The number of punctuations between two state purges determines this threshold value. Eager purge is the special case of lazy purge when the threshold is set to one. Experiments confirm that the eager strategy is suitable to minimize the join state, whereas the lazy strategy is applicable for systems with abundant memory resource.

In the second approach, Tucker et al. [104] propose *punctuation semantics* as a solution to the following problem: a join operator will need to maintain states that can grow infinitely and eventually exceed memory capacity, when continually joining multiple streams. By injecting punctuations, systems can explicitly indicate the end of a data subset, thereby enabling the safe purging of log data that will not affect future results. In this paper, the authors consider a continuous join query (CJQ) to be unsafe (and thus not permitted to run), if it requires an infinite storage. Li et al. [60] introduce the *punctuation graph* structure to analyze query safety. That is, checking whether a CJQ satisfies safety conditions under a given number of punctuation schemes, in polynomial time. To do so, they must first formally define the purgeability condition of a join operator. Then, they classify the safety verification of a CJQ into two categories: data and punctuation purgeability. The authors consider punctuation to be a special tuple that enables punctuation purging. Finally, they also propose a *chained purge* method to generalize a binary join to the *n*-way joins.

In the third approach, Li et al. [61] design a new architecture for out-of-order processing (OOP) that avoids order preservation. This is important since stream processing systems often impose an ordering of items on data streams during execution, which incurs a significant overhead when purging operator state. OOP uses punctuation or heartbeats to explicitly denote stream progress for purging operators. In addition, they introduce *joint punctuation*, a new punctuation used to reduce delay in join operators. Overall punctuation serves as a general mechanism or purge state from stateful operators [25, 60, 104].

### *3.1.4 Migrating State*

Dynamic state migration is a crucial operation in particular for stream processing systems that involve the efficient transition of state from one node to another, while preserving the operator semantics during migration. This is particularly important for operations, such as joins, aggregations, upon the addition or removal of nodes because *workloads*, *data characteristics*, and *resource availabilities* may fluctuate. Ding et al. [26] note that state migration involves two main problems: (1) *How to migrate?* That is, selecting a mechanism that reduces the overhead triggered by synchronization and delaying the production of results during migration, and (2) *What to migrate*? That is, determining the optimal task assignment that minimizes migration costs. Next, we present five approaches for migrating state.

In the first approach, Zhu et al. [121] introduce dynamic migration for continuous query plans that contain stateful operators. They propose two strategies, i.e., *moving state* and *parallel track* that exploit reusability and parallelism when seamlessly migrating continuous join query plans, while ensuring the correctness of query results. In the *moving state* strategy there are three key steps: (i) state moving, (ii) state matching, and (iii) state recomputing. Initially, the moving state step terminates the current query plan execution and purges records from intermediate queues. Then, the next step is matching and moving all records belonging to the states of the current query plan to the new query plan. This is necessary to resume the processing of the new query plan. In the *parallel track* strategy, state migrates gradually, by plugging in the new query plan and executing both query plans at the same time. Thereby, this strategy continues to produce output records throughout the migration process. When there are enough computing resources, the *moving state* strategy usually completes the migration process sooner and performs better than the *parallel track* strategy. In contrast, when resources are scarce, the *parallel track* strategy has fewer intermediate results and a higher output rate during the migration process.

In the second approach, Ding et al. [26] migrate states among nodes within a single operator. Although both *SEEP* [36] and *StreamCloud* [45] propose the idea of operator state migration, prior to Ding et al., they provide few details. In contrast, Ding et al describe algorithms that perform both *live* and *progressive* state migration. Consequently, the resulting delay prevalent in the migration process is negligible. Furthermore, they propose a (migration) task assignment algorithm that computes an optimal assignment, minimizes migration costs, and balances workloads. Moreover, they propose a new algorithm that draws on statistics from past workloads to predict future migration costs. Ding et al. criticize ChronoStream [111], which "*claims to have achieved migration with zero service disruption*," by pointing out that synchronization issues can affect the correctness of the result. To overcome this, the proposed mechanism does not migrate and execute tasks concurrently. It also ensures that all misrouted tuples are sent to their correct destinations.

In the third approach, Pietzuch et al. [91] propose a solution for migration that determines the placement locations, i.e., the selection of a physical node to manage an operator. This is indeed challenging due to variations in network and node conditions over time and the interactions among streams. In their approach, an optimizer examines the current placement of local operators and launches the migration of operators when the savings in network usage exceeds a predefined value. This *minimum migration threshold* (MMT) depends on the cost of operator migrations and maintains an operator at its current location, if the MMT is not exceeded.

In addition, they introduce *SBON* (a stream-based overlay network) that efficiently determines the placement location and reduces network utilization. The varying conditions cause SBON to re-evaluate existing placements and trigger operator migrations in new hosts, if necessary. SBON has two main components: (1) the data stream processing system, which is responsible for operations related to operator state (e.g., instantiation, migration) and data transfer, and (2) the SBON layer, which records local performance, handles the cost space, and triggers migrations.

In the fourth approach, Ottenwalder et al. [86] propose *MigCEP*, which plans migration in advance, to minimize network usage. They introduce an algorithm that generates a *Migration Plan*, i.e., a probabilistic data structure that describes future targets and times for migration. In addition, they propose another migration algorithm that minimizes both network usage and latency. It enables multiple operators to coordinate their migration (e.g., for those that may require the same mutable state) and this can further improve network utilization.

Lastly, in the fifth approach, Feng et al. [35] present two novel methods, *randomized replication representation* and an *overloaded replication scheme* to address high computational workloads (e.g., due to monitoring, migrating, replicating, and backing up states) in stateful stream processing systems. In the first method, a hashing structure, called an *MLCBF* (i.e., a Multilevel Counting Bloom Filter), replicates operators using minimal resources, to increase the performance of state migration. In addition, they use *dynamic lazy insertion*, an adaptive scheme to reduce the influence of replication, prevent the system from being overloaded, and increase cluster throughput.

### *3.1.5 Exposing State*

Exposing state in processing systems offers several advantages. For example, it: (1) enables systems to quickly recover from failures via checkpoints, (2) enables systems to efficiently reallocate stateful operators across several newly partitioned operators to provide scale out [36], and (3) facilitates integrative optimization (discussed later in Section 6). Consequently, researchers [36, 37, 38, 67, 111] have also opted to externalize state. Next, we discuss four approaches to expose state.

In the first approach, Logothetis et al. [67] propose a groupwise processing operator that considers *state* to be an input parameter. To handle state explicitly, they develop a set of flexible *primitives* for dataflow to perform large-scale data analysis and graph mining. For example, the translate operator can access state directly via a powerful *groupwise processing abstraction*, which permits users to store and access state during execution. In addition, this general abstraction supports other operations, such as *insertions, updates,* and *removals* of state. Lastly, the authors plan to develop a compiler that translates an upper-layer language into processing dataflows, to facilitate *state* access.

In the second approach, Fernandez et al. [36] seek to externalize internal operator *state*, so that stream processing systems can explicitly perform operator state management. The authors classify state into three types, namely, processing state, buffer state, and routing state. To manipulate these three types of states, they define a set of operators for state management that enables systems to *checkpoint, backup, partition,* and *restore* operator state. These primitives are the minimum set required for scale out and fault tolerance. It is possible to build more state primitives to augment the functionality. For example, the availability of abundant resources enables operator states to *merge* [45] for scale in. To deal with large state sizes, *spilling state* [63] to disk can free memory for useful computations. *Persisting* parts of an operator state into external storage enables the combination of data-at-rest and data-in-motion [5].

In the third approach, Fernandez et al. [37] make *state* explicit for imperative big data processing via the use of *SDG* (stateful dataflow graphs). Consequently, this presents a problem for big data frameworks with imperative machine learning algorithms, given that fine-grained access to large *state* is required. SDG address these challenges by efficiently translating imperative programs with large distributed state into a dataflow representation, thereby enabling low-latency iterative computation. By explicitly differentiating data from *state*, SDG use state elements, to encapsulate computation state and enable translation.

Figure 3 illustrates two distributed ways to represent an SE. One way would be to *partition* an SE and divide its data structure into disjoint parts. Another way would be to split an SE and *partially* replicate its internal data structure into multiple versions to allow for independent updates. Partitioning state across nodes can support scalability if it is possible to fully deploy the computation in parallel. On the contrary, if it is not the case, a partial SE deploys independent computations. Application semantics can then interpret these computations. The important point concerning SDG is that their tasks can directly access the distributed mutable state, allowing SDG to comprehend the semantics of stateful programs. Fernandez et al. [38] demonstrate this by developing the *JAVA2SDG* compiler to translate annotated Java programs to SDG.

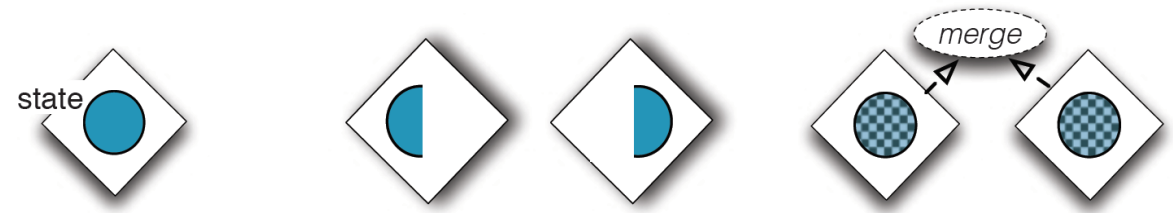

(a) State Element (b) Partitioned SE (c) Partial SE

**Figure 3.** Distributed state types in stateful dataflow graphs [37].

Lastly, in the fourth approach, ChronoStream [111] views operator *state* from two perspectives, i.e., *computation state* or *configuration state*. Computation state is a set of data structures (at the application level) that systems can directly access and conform to the user-defined processing logic. Systems hash-partition the computation state, which is kept in an operator, into an array of fine-grained computation slices. To enable load balancing, slices are distributed equally among resource containers. Every subset of input data corresponds to an independent slice that generates a corresponding output stream. Configuration state is a collection of states (at the container level), which is used to maintain runtime parameters. This state is associated with each resource container and its contents differ among containers. The configuration state associated with each container comprises three components: (1) an *input routing table*, to deliver input events to corresponding slices, (2) an *output routing table*, to direct output events to a resource container associated with a downstream operator, and (3) a *thread-control table*, to preserve the thread schedule (at the operating system level) and compute the upper-layer slices. Generally, configuration state plays a role as the intermediate connection between parallelism at the application level and local multithreads at the operating system level.

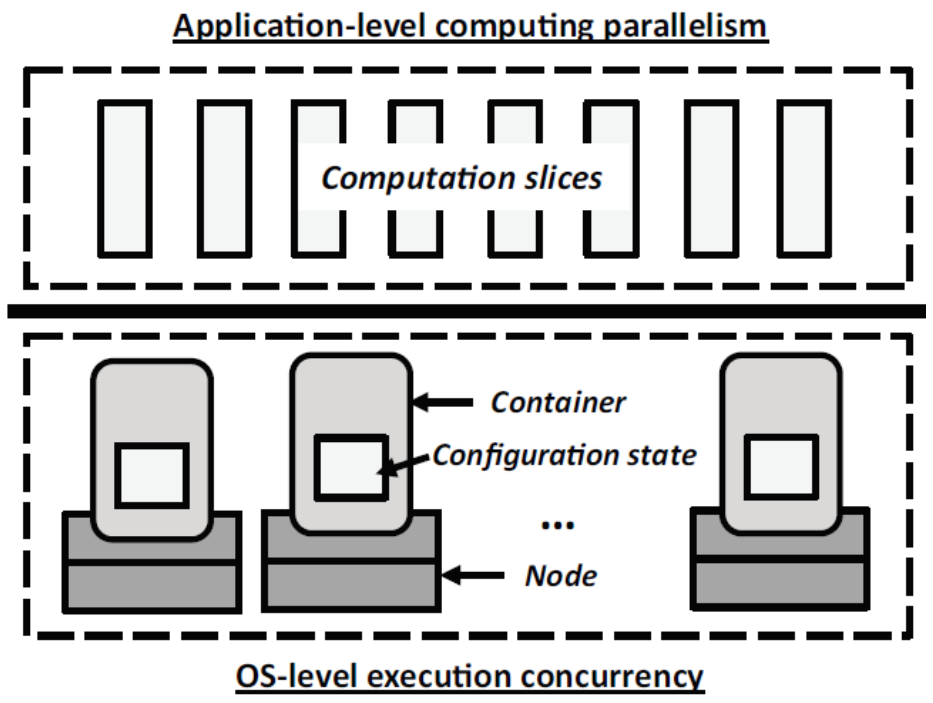

**Figure 4.** An internal state management abstraction design [111].

Figure 4 illustrates the relation and order among computation states, configuration states, resource containers, and the computing nodes in an operator. Using the concept of slices, ChronoStream supports horizontal and vertical elasticity by scaling the underlying computing nodes logically and managing the configuration states associated with these nodes rather than handling the computation states at the application level.

## 3.2 State Sharing

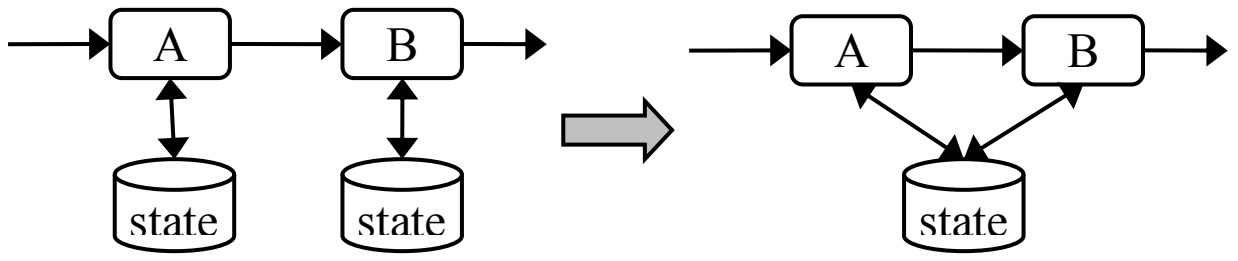

**Figure 5**. State sharing.

State sharing (cf. Fig. 5) denotes using state for several operations during data flow processing. This is desirable in many instances. For example, it can reduce data transmission over networks and thus reduce latency. Table 2 provides a glimpse into four systems, where state is shared. These are presented next.

State sharing facilitates optimizing stream processing systems. For example, Hirzel et al. [48] examine a streaming application that continuously calculates statistics (e.g., average stock price) for different time windows (e.g., hours, days). Since these operations differ only on the *time granularity* (e.g., hours vs. days), then it is natural to share the aggregation window. By doing so, this will increase resource utilization (e.g., memory) efficiency among operations. However, sharing state can lead to some inherent problems, such as access conflicts, consistency issues, or deadlocks. Therefore, Hirzel et al. point out three safety conditions requirements. First, *ensuring visibility* can make state visible and accessible to all operators. Second, *prevention of race conditions* can assure state is immutable and/or that synchronization among processes is properly set. Lastly, *safe management of memory* can prevent the early release of memory or uncontrollable expansion, which could lead to memory leaks.

| Common Characteristics | System | Main Mechanism | Objective |
|---|---|---|---|
| They avoid computation or transmission redundancies to achieve higher performance. | [48] | focus on safety conditions | discuss multiple forms of sharing |
| | [59] | in-network query processing and multi-subscription optimization | eliminate unnecessary computation |
| | [68] | use data structures at the language level | share state across operators |
| | [75] | ensure both correctness and ACID guarantees | target transaction processing |

**Table 2**. A characterization of state sharing methods.

In their paper, Hirzel et al. discuss three forms of state sharing. The first form involves *shared operator state* [15], where state can be arbitrarily complex. In this form, synchronization and memory management present key challenges. Indeed, sharing memory may introduce conflicts, which are often resolved using mutual-exclusion locks. However, when conflicts are rare, this approach is cost prohibitive (e.g., when performing concurrency handling). Therefore, an alternative approach [15] uses *software transactional memory* to manage data sharing. The second form entails *shared windows* [9, 44] that enable multiple consumers to utilize the same window. Window sharing is indeed one of the simplest cases of state sharing [44]. For example, the continuous query language (CQL) implements windows by using non-shared arrays of pointers to reference shared data. This model of many-to-one pointer reference can allow many windows and event queues [9] to access a single data item. Lastly, the third form encompasses *shared queue* [97]. Here, the simultaneous access of both producer and consumer to a single element (i.e., the producer writes a new item and the consumer concurrently reads an old item) can lead to conflicts. To guarantee synchronization and preserve concurrency, queues must be able to buffer two data items at a minimum.

In their paper [59], Kuntschke et al. recognize instances of computational inefficiencies in large-scale data processing that can be eliminated by sharing state. Examples of these include the unnecessary execution of operators and data transfers, among other redundancies. By sharing data streams, we avoid redundant transmissions and save network bandwidth. Another benefit of discarding unnecessary computation is reducing the execution time, by sharing previously computed results and early filtering and aggregation (e.g., the combine function in MapReduce). They propose two optimization techniques: *in-network query processing*, which distributes and performs (newly registered) continuous queries and *multi-subscription optimization*, which enables the reuse and sharing of generated data streams.

In their paper [68], Losa et al. propose *CAPSULE*, a language and system that supports *state sharing* across operators, using a less structured method than point-to-point dataflows. It shares variables (a.k.a. states) using a data structure at the language level. Besides supporting the efficient sharing of state in distributed stream processing systems, CAPSULE provides three features. That is, (i) *custom code generation*, to produce shared variable servers that fit a given scenario based on runtime information and configuration parameters, (ii) *composability*, to achieve suitable levels of scalability, fault-tolerance, and efficiency using shared variable servers, and (iii) *extensibility*, to support, for example, additional protocols, transport mechanisms, and caching methods, using simple interfaces.

In their paper [75], Meehan et al. introduce S-Store, a system designed to maintain correctness and *ACID* guarantees (i.e., atomicity, consistency, isolation, and durability) that are essential to handle *shared mutable state*. By employing shared state, the system achieves high throughput and consistency for both transaction processing and stream processing applications. In this context, the proper coordination and sharing among successive executions of a window state differ from other sorts of state (e.g., where state is privately shared with other transactions). In this way, S-Store achieves low latency with correctness in stream processing and high performance with ACID guarantees in transaction processing. Tatbul et al. [101] further explore correctness criteria, including ACID guarantees, ordered execution guarantees, and *exactly-once* processing guarantees. To support these three-complementary correctness guarantees, S-Store provides efficient scheduling and recovery mechanisms. Although Naiad, SEEP, and Samza all view state as mutable, they do not inherently support transactional access to shared state. Thus, Meehan et al. [75] show that the consistency guarantees offered by S-Store are better than the consistency guarantees offered by Naiad, SEEP, and Samza.

## 3.3 Incremental Maintenance

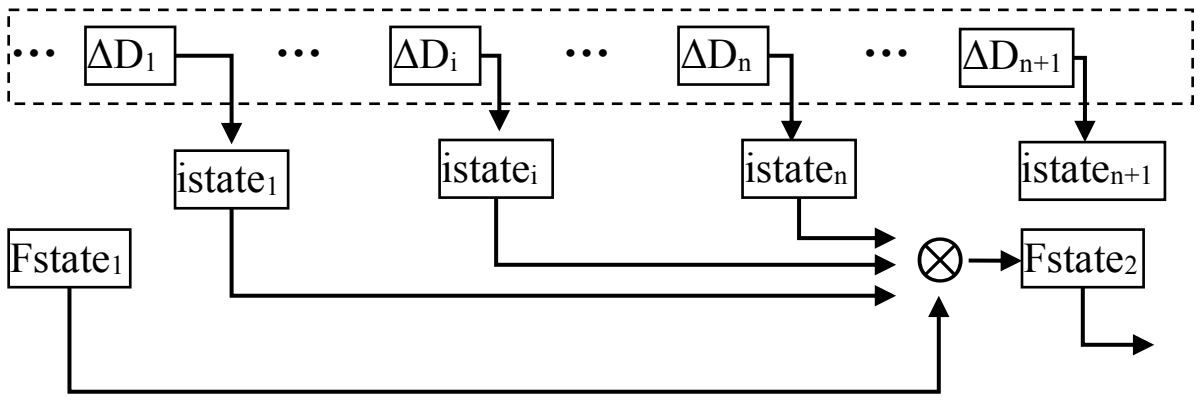

**Figure 6**. Incremental maintenance of state.

Researchers have sought to *reduce incremental checkpointing overhead* [82, 96] or *maintain state incrementally* [33, 54, 55, 56, 74, 84, 85] to cope with frequent data updates and avoid costly full state updates. By generating delta values (cf. Fig. 6), they can all update persisted state more efficiently, whenever inputs vary marginally, and avoid recomputing from scratch. Table 3 provides a glimpse into seven approaches that maintain state incrementally. Next, we elaborate on these approaches.

The first approach due to McSherry et al. [74] presents *differential computation*, which generalizes existing methods for incremental computation with continuously changing input data. Their method differs from traditional incremental computations by supporting *arbitrarily nested* iterative computations. Akin to the Naiad system, the key innovations come from two factors. First, changes in state adhere to a *partially ordered* sequence, instead of a totally ordered one, which conforms to incremental computation. Second, an indexed data-structure maintains a set of updates that is essential to rebuild the state. This second feature is different from the other incremental systems, in that updates are usually discarded after being merged with the current state snapshot.

The second approach due to Koch [54], employs monoid algebra to address the *incremental view maintenance (IVM)* problem and extends an algebraic structure of a *ring of databases* to form a powerful aggregate query calculus. This calculus inherits the key properties of *rings*, such as distributivity and the existence of an additive inverse. Thereby, this makes the calculus closed under a universal difference operator that expresses the delta queries of the IVM. These key properties provide the basis for delta processing and incremental evaluation. The multi-layered IVM scheme can maintain a view (using a hierarchy of auxiliary materialized views) and refresh it, whenever there are updates. Furthermore, their findings lay a foundation for subsequent research [5, 55, 84, 85] in incremental state maintenance.

The third approach due to Fegaras [33], introduces a prototype, called *MRQL Streaming*, that returns (at each time interval) continuous answers, by merging the last materialized state and the delta result of the most recent data batches. The novelty of this approach comes from algebraic transformation rules that convert queries to homomorphisms. MRQL Streaming decomposes a non-homomorphic streaming query $q(S)$ into two functions, $a$ and $h$, such that $q(S) = a(h(S))$, where $h$ is a homomorphism (i.e., $h(S + \Delta S) = h(S) \otimes h(\Delta S)$) and $a$ is a non-homomorphic component of the query that forms the answer function. Accordingly, state stores the result of the incremental calculation $h$, using the current state value to compute the next $h$ value (i.e., $state = state \otimes h(\Delta S)$). Initially, state is either empty or set to $h(S)$, if there are initial streams. Then, at every interval, $\Delta t$, the answer to the query is computed from the state that is equal to $h(S + \Delta S)$.

| Common Characteristics | System | Main Mechanism | Targeted Computation |
|---|---|---|---|
| N/A, only one system | [74] | a partially ordered sequence, preserves a set of updates to rebuild state | arbitrarily nested iterative computations |
| Monoid algebra | [54] | algebraic rings in databases | aggregate query |
| | [33] | algebraic transformations with lineage tracking and homomorphisms | iterative and nested queries, group-by with aggregation, equi-joins |
| Delta computations based on algebra | [5] | recursive finite differencing technique | general incremental view maintenance |
| | [84] | matrix factorization | linear algebra program iterations in machine learning |
| | [85] | derive delta programs to capture changes in the result | queries with nested aggregates |
| | [56] | nested relational calculus | bag computing |

**Table 3**. A characterization of incremental state maintenance methods.

The fourth approach due to Ahmad et al. [5], introduces a recursive, finite differencing technique, called *viewlet transforms*, that unifies historical and current data. Their technique materializes a query and its corresponding views, which support the mutual incremental maintenance, thereby, reducing the overall view maintenance cost. Similarly, Koch et al. [55] fully describe and experimentally evaluate the performance of the DBToaster system, using the ring theory. DBToaster can continuously update materialized views, despite frequent data changes, using an aggressive compilation technique or a recursive finite differencing technique.

The fifth approach due to Nikolic et al. [84], introduces the *LINVIEW* framework and the concept of *deltas*, which captures changes to linear algebra programs (LAP) and highlights the use of IVM in LAP involving iterations in machine learning. Linear algebra operations can trigger a ripple effect (e.g., small input changes can propagate and affect intermediate results and the final view). This can negatively affect the performance of IVM upon re-evaluation. To mitigate this problem, LINVIEW employs matrix factorization methods to enable IVM to be suitable and less expensive than recomputing from scratch.

The sixth approach due to Nikolic et al. [85], generalizes the results of Koch et al. and presents recursive and incremental techniques to handle queries containing nested aggregates. They compare the performance between tuple and batch incremental updates to identify scenarios when batch processing can substantially improve the efficiency of IVM. Their experimental findings show that single-tuple execution outperforms generic batch processing in many situations, thus contradicting the belief that batch processing outperforms single-tuple processing [87].

Lastly, the seventh approach due to Koch et al. [56], provides an efficient solution to incrementally compute the positive *nested*

*relational calculus* (NRC+) on bags. They develop a cost model for NRC+ operators that enables them to calculate the cost of delta computations. A query can be considered *efficiently incrementalizable* if the cost of its delta is strictly lower than that of recomputation from scratch. A large part of NRC+, called IncNRC+, which satisfies the efficient incrementalization condition is translated from NRC+ without losing its semantics.

## 3.4 Load Balancing and Elasticity

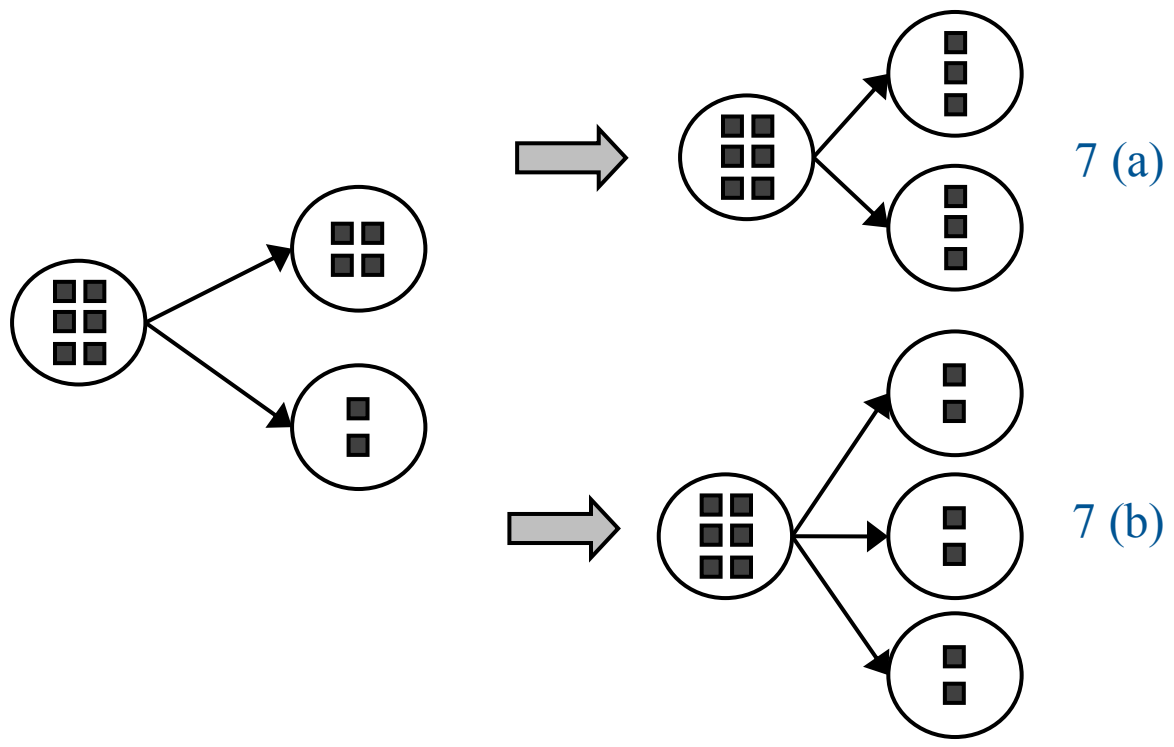

**Figure 7**. State in load balancing and elasticity.

System workloads are dynamic and when demands increase these are typically managed via the concept of load balancing or elasticity. *Load balancing* characterizes a computing system's ability to redistribute its workload across computing resources, particularly, when some nodes have heavier loads than others. For example, when the workload in a node increases, it can be redistributed to another node to ensure workload balance, as depicted in Figure 7 (a). *Elasticity* characterizes a computing system's ability to provide additional computing resources in light of increasing workloads. For example, with increasing workloads, we can allocate additional resources (i.e., nodes) to share the workload, as depicted in Figure 7 (b). Although handling elasticity and load balancing in stateless operators is straightforward, it is challenging for stateful operators due to the complexity in managing state. Today's data-parallel computation frameworks handle *elasticity* by maintaining and migrating state, while jobs are actively running.

To migrate state, the number of parallel channels need to dynamically adapt (i.e., nodes are added or removed) at runtime to match the computing resources and workload availability, which may unexpectedly fluctuate. Thus, in the presence of workload skew, the states of heavy burdened nodes are repartitioned and reallocated (cf. Fig. 7) to nodes that are not burdened. Similarly, when resources are scarce, the states of tasks that are affected (e.g., job partitions) need to be reallocated. Hence, we require partitioning methods that enable systems to scale and achieve workload balance. These mainly fall into four types, i.e., hash-based, partial key based, state migration [111], and executor-centric [128]. Table 4 characterizes several systems by their respective partitioning type. Next, we examine varying systems that fall under these four partitioning types.

Dataflow scalability in streaming systems is limited by stateful operators. In order for these operators to scale, they will need to be partitioned (e.g., across a shared-nothing platform). However, over time, this will lead to load unbalancing. To resolve this problem, Shah et al. [98] propose *Flux*, a dataflow operator that encapsulates adaptive state partitioning and dataflow routing. Placed between producer and consumer stages in pipelined dataflows, Flux repartitions stateful operators transparently, without interrupting the pipeline under execution. Flux provides two mechanisms to adapt to both short term and long-term imbalances. In the short-term case, Flux utilizes a buffer and a reordering mechanism to adjust local imbalances. In the long-term case, Flux detects imbalances across the entire cluster and allows state repartitioning in lookup-based operators to manage the problem.

| Partitioning Type | System | Main Focus | Objective |
|---|---|---|---|
| Hash based | [98] | state partitioning and dataflow routing | distribute workload uniformly across computing nodes |
| Partial-key based | [43] | partition functions | |
| | [126] | add aggregation cost to model | |
| | [127] | key splitting and local load estimation | |
| | [139] | associates a key to more than two possible nodes | |
| Executor-centric | [128] | elastic executors + model-based scheduler | |
| Migration-based | [111] | transactional migration protocol and thread-to-slice mapping | |

**Table 4**. A characterization of partitioning schemes.

Gedik et al. [43] devise new partitioning functions to redistribute skewed workloads, which trigger imbalances (e.g., memory usage, computation, communication costs across parallel channels). In addition, they introduce several desirable properties that these functions must meet. These properties include: (1) *balance properties* (e.g., memory, communication and processing balance), (2) *structural properties* (e.g., fast lookup, compactness), and (3) *adaptation properties* (e.g., minimal migration, fast computation). Experiments show that the proposed partitioning functions possess these desirable properties over a variety of workloads and thus provide better load balance than uniform and consistent hashing. These functions are especially effective for workloads with large key domains (i.e., the cardinality of the partitioning key). In this case, they can efficiently balance communication costs, computation costs, and memory load, yet still ensure low migration overhead despite workload skew.

Nasir et al. [127] propose a stream partitioning scheme, called partial key grouping (PKG), to partition the load in distributed stream processing systems. PKG includes two main techniques, i.e., *key splitting* and *local load estimation*. The key splitting technique is based on the "*power of two choices*" principle, in which the system selects two nodes uniformly at random and delivers the streaming element into the one that has the least load. In the local load estimation technique, each source operator maintains a local load-estimate vector, which is calculated by using only local information about the portion of stream sent by each source. Experiments show that PKG achieves better load balancing than standard hashing. However, in the case of large deployments, solely having these two choices is insufficient, since skew is inversely proportional to the size of the deployment. Therefore, to remedy this, Nasir et al. [139] propose two streaming algorithms, called *D-Choices* and *W-Choices*, to enable load balancing in large deployments. Experiments show that these two algorithms achieve very low imbalance (i.e., smaller than 0.1%) in large deployments.

Katsipoulakis et al. [126] uses a partitioning algorithm to ship records to computing nodes. However, they integrate the *aggregation cost* into the cost model to improve performance. In this model, the aggregation combines all of the partial results corresponding to the partitioned operations produced by computing nodes. While previous works focus only on load imbalance, combining load imbalance and aggregation cost improves the balance among computing nodes, and therefore reduces the overall latency of the system. This combined method achieves the best performance over competing methods when the number of groups (in *group-by* operators) is large.

Wang et al. [128] propose the Elasticutor framework to achieve elasticity by an executor-centric method. Here, executors are parallel execution instances, and play the role of building blocks for elasticity. Instead of partitioning the key space of an operator *dynamically* as in key partitioning methods [126, 127], this method partitions the key space *statically*, but allocates CPU cores to executors *dynamically*. Elasticutor applies optimization at two levels: (1) a scheduler that assigns CPU cores to executors at the global level, and (2) a subsystem that allocates workloads to these cores at the executor level.

*ChronoStream* [111] takes a different approach to address the load balancing and elasticity problem. By treating the internal state as a built-in component, ChronoStream achieves flexible scalability. That includes *horizontal elasticity*, where resources vary in all of the computing nodes and *vertical elasticity*, where resources vary at a single node. Consequently, this enables ChronoStream to efficiently manage both workload fluctuation and dynamic resource reclamation. For horizontal elasticity, transparent workload re-allocation is achieved using a lightweight *transactional migration protocol* based on the reconstruction of state at the stage-level. To support vertical elasticity, ChronoStream provides fine-grained runtime resource allocation that maps an OS-level thread to many application-level computation slices. A *thread-control table* stored in the configuration state can be used to record this thread-to-slice mapping. To scale vertically, ChronoStream utilizes this table to reschedule the computation. At any time during the execution, the workload in each thread can be dynamically reorganized to rebalance the load (i.e., dynamic re-partitioning).

## 3.5 Performance

Managing state can incur significant overhead, including increased processing latency and recovery time. Hence, varying performance optimization techniques have been proposed to reduce the overhead. For example, setting the intervals among checkpoints when storing and replicating state for fault-tolerance purposes appropriately can substantially reduce the execution time of an iterative algorithm [94]. The state checkpoint placement problem has been shown to be NP-complete [13]. The overhead and complexity associated with state management approaches vary widely. Next, we discuss some issues related to the performance of state management, such as the *impact of frequent checkpointing* (determined by checkpoint interval calculations), the *complexity of optimal state placement*, and the *complexity of optimal state assignment*.

### *3.5.1 Impact of Frequent Checkpointing*

In practice, heuristics are often used to decide when to checkpoint state (e.g., periodic or aperiodic checkpointing). Periodic checkpointing enables systems to quickly recover from failure. However, systems will expend resources and time that could be better used elsewhere. In contrast, aperiodic checkpointing leads to longer failure recovery times. Thus, in recent years, systems researchers [36, 82, 94] have focused on determining an optimal checkpointing frequency.

Naksinehaboon et al. [82] investigate the optimal placement of checkpoints to minimize the total overhead, i.e., both the *rollback recovery* and *checkpointing* overhead. By employing a checkpointing frequency function, they can derive an optimal checkpointing interval based on a user-provided failure probability distribution.

Fernandez et al. [36] measure processing latency and demonstrate that aperiodic checkpointing would generate varying latencies. Their method reveals that wider intervals have less impact on data processing, but lengthen the failure recovery time. Instead, they propose setting the checkpointing interval, according to the estimated failure frequency and the query performance requirements.

Sayed et al. [94] evaluate the impact of checkpointing intervals across methods. They critique ad-hoc periodic checkpointing rules, such as *checkpointing every 30 minutes*. They observe that the model due to Young [114] achieves near optimal performance and is applicable in practice. They further investigate more advanced methods that dynamically change the check-pointing interval. Their findings show that these methods significantly improve over Young's model for only a small subset of systems.

### *3.5.2 Complexity of Optimal State Placement*

Determining when to effectively place checkpoints is yet another challenging problem. Researchers [13, 93] formally prove that this problem is NP-complete and propose approximation algorithms to solve this problem in polynomial time.

Robert et al. [93] focus on the *complexity of computational workflow scheduling* with failures that follow an exponential distribution. They aim to optimize the expected processing time, processing schedule of independent tasks, and checkpointing time, which are combinatorial problems. They prove that this optimization problem is *strongly NP-complete* and propose a dynamic programming algorithm that runs in polynomial time.

Bouguerra et al. [13] examine the *computational complexity of checkpoint scheduling* with failures that follow arbitrary probability distributions. They note that both *costs among checkpoints* as well as the *processing time for data blocks* vary. Therefore, they develop a new complexity analysis to exploit relationships among *failure probabilities*, *checkpoint overhead,* and a *computational model*. Additionally, they introduce a new mathematical formulation to optimize checkpoint scheduling in parallel applications. They prove that checkpoint scheduling is NP-complete and propose a dynamic programming algorithm to determine the optimal times for checkpointing.

### *3.5.3 Complexity of Optimal State Assignment*

Determining an effective strategy to partition tasks efficiently is a challenging problem, given that the size of the search space is exponential. Ding et al. [26] calculate the optimal task assignment to minimize state migration costs (i.e., the total storage size of all the operator states transferred among nodes) and meet load balancing conditions. Adhering to their notation, let the output of partitioning function $f$ to input record $r$ be an integer $f(r)$, with $1 \leq f(r) \leq m$. Each node $N_i$ $(1 \leq i \leq n)$ is assigned an interval $I_i = [I_i.lb, I_i.ub)$, $1 \leq lb_i \leq ub_i \leq m$, called the *task interval* of $N_i$. Given a threshold $\tau$, a task assignment is considered to be load balancing if and only if the workload $W_i$ for each node $N_i$ satisfies this condition $W_i \leq (1+\tau)W/n$. In other words, this condition means that each node does not have too high workload when comparing to the average value of the perfect case where every node shares exactly the same amount of work *W/n*. The optimal task assignment includes two consecutive steps: dividing all tasks into *n'* separate task intervals, and then allocating these task intervals to *n'* different nodes.

To address the task partitioning problem, the researchers split it into numerous sub-problems, then solve each sub-problem using *Simple_SSM*, a proposed basic solution with $O(m^2n^2n'^2)$ possible sub-problems. *Simple_SSM* incurs a space complexity of $O(m^2n^2n'^2)$ and time complexity of $O(m^3n^3n'^2)$. To improve upon this, they propose another solution that exploits optimizations and gradually improves the space and time complexity over time. The best solution uses only $O(mn')$ space and $O(m^2n')$ time, which is a significant reduction over the basic solution.

# 4. Applications of State

This section presents three applications of state. This includes the use of state in stateful computation, iterative processing, and fault tolerance.

## 4.1 Stateful Computation

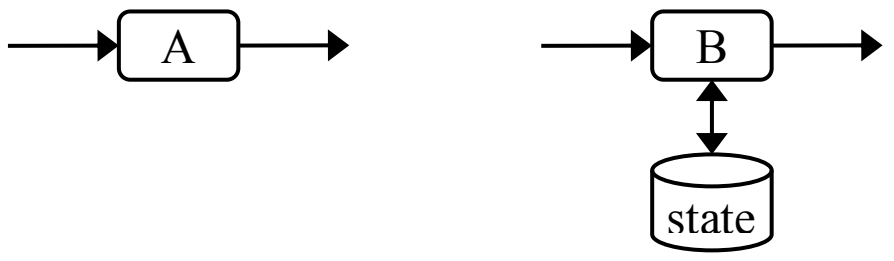

**Figure 8**. State in stateful computation.

Naturally, *state* serves to enable stateful computations during data stream processing. Computation on records of a data stream can either be *stateless* or *stateful*. In stateless operators (e.g., filtering), there is no record of previous computations. Instead, each computation is purely functional, handled entirely based on the current input. By definition, stateful operators (e.g., aggregations over time windows or some other stream discretization) interact with earlier computations or data observed in the recent past. Thus, since *state* represents prior computational results or previously seen data, it must be persisted (cf. Fig. 8) for subsequent use. This is evident in today's popular data stream processing frameworks, such as Flink, Spark, Storm, Storm + Trident, and Heron, each of which supports stateful operators.

Despite commonalities among frameworks, there are contrasting views on how to best implement state. For example, early versions of Storm focused on stateless processing and required state management at the application level. Storm + Trident (an extension of Storm) enables state management via an API. Samza manages large states using a local database to enable persistence. Spark Streaming enables state computation via DStream (i.e., discretized streams). Finally, Flink treats state as a first-class citizen, which eases stateful application development. The implementation of state across four frameworks is discussed in greater detail in Section 11. Table 5 captures the characteristics of stateful computation methods across four systems. Next, we discuss representative papers centered on stateful computation.

In the late 2000s, bulk data processing systems, like MapReduce were growing in popularity. However, they were criticized for not offering data indexing, which as a form of efficient state access could conceivably increase performance. These findings lead Logothetis et al. [66] to devise a data indexing scheme to support *stateful groupwise processing*. They observe that by offering access to persistent state, operations, such as reduce, could cope with data updates and circumvent the need to recompute from scratch. Additionally, that indexing can avoid expensive sequential scans and grant groupwise processing random access to state.

Logothetis et al. [67] discuss two (suboptimal) solutions for stateful bulk processing. One solution requires running the entire dataflow once again, whenever new data arrives. In contrast, the other solution requires programmers to employ data-parallel programs, to incorporate and use state. However, due to limitations in frameworks, such as MapReduce, this will be difficult. Instead, they propose an alternative approach by treating state as an explicit input that can *store* and *retrieve* as new data arrives. Moreover, by employing a stateful groupwise operator (i.e., *translate*), data movement is minimized and state is smoothly integrated into a data-parallel processing framework.

| Common Characteristics | System | Main Mechanism | Objective |
|---|---|---|---|
| Batch processing | [66] | indexing | avoid a sequential scan |
| | [67] | state as explicit input | minimize data movement |
| Stream processing | [43] | partitioned stateful operators | balance the load |
| | [72] | parallel patterns | increase parallelism |

**Table 5**. A characterization of stateful computation methods.

Gedik et al. [43] exploit *partitioning functions* for stateful data parallelism in stream processing systems to improve application throughput. They note that partitioned stateful operators (PSO), such as streaming aggregation, one-way join, and progressive sort are well-suited for data parallelism and demonstrate that these can hold state on partitioning-key defined sub-streams. Furthermore, they indicate that for PSO, hash functions must be employed "*to ensure that tuples with the same partitioning key value are routed to the same parallel channel*." In conclusion, they reiterate that partitioning functions enable adequate memory load balance, communication, and computation, while concurrently maintaining the migration overhead low under a variety of workloads.

Matteis et al. [72] address parallelism challenges involving stateful operators arising in modern stream processing engines (e.g., Spark Streaming, Storm) by *algorithmic skeletons*. Algorithmic skeletons (a.k.a. parallelism patterns) are a high-level parallel programming model for parallel and distributed computing. They are useful in hiding the complexity parallel and distributed applications. They present four parallel patterns for window-based stateful operators on data streams: *window farming*, *key partitioning*, *pane farming*, and *window partitioning*.

The *window farming* pattern (WFP) applies each computation (e.g., a function) to a window and the corresponding results will be independent of one another. The *key partitioning* pattern extends the WFP by adding a constrained assignment policy. In this policy, the same worker processes windows originating from the common sub-stream sequentially, however, this limits the parallelism.

The *pane farming* (PF) pattern splits each window into non-overlapping partitions called panes. This fine-grained division increases throughput and decreases latency by sharing the results of overlapping panes. Finally, the *window partitioning* pattern requires multiple workers to process each individual window. Akin to PF, this pattern improves throughput and reduces latency.

However, this latency reduction depends on the total number of workers, in contrast to the pane farming pattern, which does not.

## 4.2 Iterative Processing

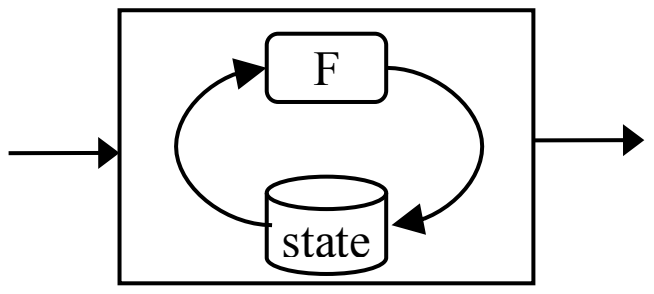

**Figure 9**. State in iterative processing.

State can be used to efficiently enable iterative processing (IP) in big data frameworks (BDF) (cf. Fig. 9). IP continuously applies a user-defined function (often called a *step function*) to a data collection until a convergence criterion (e.g., a fixed point, a fixed number of iterations) is met. This type of operation is of paramount importance for large-scale data analysis since most machine learning and graph mining algorithms are iterative in nature. Yet, they are ill-suited for BDF, such as MapReduce [24], since they incur a large overhead, in particular for many graph or social network analysis algorithms. These often times needlessly reload and reprocess data during iterations; even though they leave large parts of the data unchanged [124]. Additionally, each iteration is executed as a separate job [95], which prevents optimizations across iterations. These drawbacks lead to the development of iterative mechanisms and their integration into data-parallel processing systems [31, 32]. In his vision paper [70], Markl affirms that the native support of *state* in iterative data analysis programs is a key design for future platforms.

Iterative computations come in two varieties, namely, *bulk* and *incremental*. In bulk iterations, each step produces an entirely different intermediate computation in contrast to the (final) result. Examples of bulk iteration include machine learning algorithms, such as batch gradient descent [109] and distributed stochastic gradient descent [122]. In incremental iterations, the result of a current iteration (at time step *i*) slightly differs from the result of the previous iteration (at time step *i-1*). As discussed in [95], the elements of the intermediate computations exhibit "*sparse computational dependencies*." That is, changes in one element solely affect a few other elements. For example, in the connected components algorithm, an update to a single vertex impacts only its surrounding neighbors. Table 6 lists varying systems classified by their common characteristics. Next, we discuss four papers/systems and their proposed approaches for iterative processing involving state.

The first approach, due to Ewen et al. [31], overcomes some performance issues in existing dataflow systems, which treat incremental iterations as bulk iterations. As a result, some iterative algorithms perform poorly. To resolve this, the authors devise a method that integrates incremental iterations into parallel dataflow systems, by exploiting *sparse computational dependencies* that are intrinsic in many iterative algorithms. Rather than creating a specialized system, their method facilitates expressing analytical pipelines in a unified manner and disregards the need for an orchestration framework. As a proof-of-concept, the authors [32] illustrate the implementation, compilation, optimization, and execution of iterative algorithms in *Stratosphere*.

The second approach due to Fegaras [33], called *MRQL Streaming*, improves iterative processing performance over the two earlier approaches. It relies on two techniques, namely, *lineage tracking* [12] and *homomorphisms*, to reduce the state size. In the lineage tracking technique, attributes in *join* and *group-by* clauses are moved to query outputs, to establish connections between the input data and query results. In contrast, the homomorphism-based technique combines the current state value with new input data to generate new output. To apply these two methods, MRQL Streaming automatically converts a SQL query to an incremental, distributed program that runs on a stream processing engine. Then, it derives incremental programs by storing a small state during the query evaluation process and using a novel incremental evaluation technique that merges the current state value and the latest data.

The third approach, due to Schelter et al. [95], utilizes state to address fault-recovery during the iterative processing of fixpoint algorithms, which are common in machine learning. In their paper, the authors introduce a mechanism based on the principle of algorithmic compensations to achieve optimistic recovery. *Algorithmic compensations* concern the exploitation of a fixpoint algorithm property, namely, the ability to converge to the solution from several intermediate consistent states. *Optimistic recovery* concerns resuming computation from the latest iteration, in contrast to rollback recovery, where computation starts from scratch. Using their ideas, the authors are able to rebuild state, using a user-defined, algorithm-dependent *compensation function*. Furthermore, their approach outperforms rollback recovery methods, since state checkpointing occurs in the background, independent of and not interfering with the processing of data. Additionally, they show how their method can be employed in three areas: *factorizing matrices*, *performing linking and centrality computations* in networks, and *identifying paths in graphs*. Lastly, Dudoladov et al. [27] demonstrate the efficiency of the optimistic recovery mechanism for both the Connected Components and PageRank algorithms in Apache Flink.

| Common Characteristics | System | Main Mechanism | Objective |
|---|---|---|---|
| Integrate with incrementalization | [31] | exploit sparse computational dependencies | improve performance |
| | [33] | lineage tracking and homomorphism | reduce state size |
| Occurs transparently in the background | [95] | use algorithmic compensations | achieve fast recovery |
| | [112] | unblocking mechanism for checkpointing | |

**Table 6**. A characterization of iterative processing methods.

The fourth approach, due to Xu et al. [112], introduces the concepts of *head* and *tail* state checkpointing, to lower checkpointing costs and reduce failure recovery time. Head (tail) checkpointing writes checkpoints at the beginning (end) of a step (i.e., each iteration in the iterative computation). In their approach, they use an *unblocking* mechanism to write checkpoints, transparently in the background without requiring the program to interrupt. This avoids the overhead associated with delayed execution at checkpoint creation time. By injecting checkpoints directly into dataflows, this method takes advantage of both *low-latency execution* (by disregarding pipeline process interrupts), and the *seamless integration into existing systems*. Furthermore, the use of local log files on each node circumvents the need to recompute from scratch upon failure and yields a faster (or *confined*) *recovery*.

## 4.3 Fault Tolerance

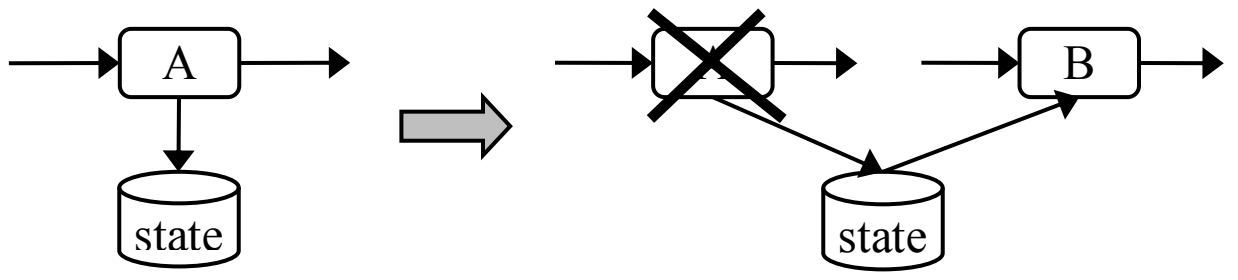

**Figure 10**. State in fault tolerance.

State can be used to enable failure recovery ~~fault tolerance~~ and thereby facilitate fault tolerance. It is persisted in reliable storage and updated periodically. When failure occurs, big data processing systems restore the state to another node, thereby, recovering the computation from the last checkpoint (cf. Fig. 10). Fault tolerance, in general, requires redundancy, which can be achieved in several ways. One approach enables the redundant storage (or replication) of computations. A second approach enables the redundant storage of the computational logic, which involves a variant of state called *lineage* (e.g., prevalent in Spark). Alternatively, a third approach employs redundant computation [95], which exploits algorithmic properties and does not use state.

According to Hwang et al. [49], there are three fault-tolerance mechanisms: *passive standby*, *active standby*, and *upstream backup*. In the case of passive standby, only the modified part of the state is backed up periodically. In the case of active standby, redundant execution enables each backup server to receive and process the same input from upstream servers, in parallel, as its primary server. Lastly, in the case of upstream backup, each primary server retains its output, while the backup is still inactive. If a primary server fails, the backup restores the primary server's state by reprocessing tuples stored at upstream servers.

Each method has its own advantages, in terms of network usage, recovery latency, recovery semantics, and system performance [49]. Most researchers prefer passive standby (or checkpointing), to achieve fault-tolerance because it is effective in addressing more configuration and workload needs than the alternative approaches [50]. Additionally, this method reduces the overall recovery overhead, since each checkpoint can be restored in parallel. Orthogonal to the taxonomy of Hwang et al. [49], we classify fault-tolerance methods into three key categories, i.e., independent, dependent, and incremental. These categories are determined via the state handling approach employed, as a classification criterion.

### *4.3.1 Independent Checkpointing*

In the research literature, there are two types of (node) failures, namely, *independent* failures and *correlated* failures. The assumption is that failures are either independent of one another or occur simultaneously, i.e., correlated. Table 7 summarizes varying independent checkpointing methods by their shared characteristics. Next, we discuss three methods for independent checkpointing.

Hwang et al. [50] introduce the concept of a *maximal connected subgraph,* which they regard as an atomic (i.e., a high-availability or HA) unit for independent checkpointing. These units can be checkpointed onto independent servers at varying times since they have no interdependencies and thus avoid inconsistent backup checkpoints. Consequently, spreading out independent checkpoints to multiple servers can reduce the checkpointing overhead. Comparably, Kwon et al. [57] split state into partitions that can independently checkpoint states, while ensuring consistency in the event of node failures.

Sebepou et al. [96] produce independent partial checkpoints asynchronously, by splitting operator state into disparate parts. Represented as control tuples, these independent checkpoints contain the partial state of an operator and combined with normal tuples, which contain the actual data in operator output queues. As a consequence, this enables us to use a single persistent architecture for both the operator and output queues. This approach follows the upstream backup mechanism by persisting the output queue to stable storage. In the event of node failures, an operator's input queue can be rebuilt by fetching tuples from an upstream operator's output queue.

| Common Characteristics | System | Main Mechanism | Objective |
|---|---|---|---|
| Each performs independent checkpointing | [50] | maximal connected subgraphs (MCS) | checkpoint MCS independent of one another |
| | [57] | checkpoint asynchronously | partitioning state, ensure consistency |
| | [96] | use control tuples to represent the partial state of operators | use single persistence for operator and output queues |

**Table 7**. A characterization of independent checkpointing methods.

### *4.3.2 Correlated Checkpointing*

Correlated failure events involve the simultaneous failure of multiple nodes. They generally occur, whenever switches, routers, or electrical power fail. Indeed, when failures occur, varying coping strategies [20, 46, 58, 100, 105, 108] have been proposed. Table 7 summarizes varying systems by shared characteristics. Next, we dive into seven approaches for correlated checkpointing.

| Common Characteristics | System | Main Mechanism | Objective |
|---|---|---|---|
| Uninterrupted processing | [132,133] | replication and upstream backup | availability & consistency |
| | [20] | adapts to failures | improve scalability |
| | [108] | injects tokens into streams | not interrupt operator |
| Employ multiple checkpointing methods | [100] | combines passive and active checkpoints | save resources (storage) |
| | [105] | uses varying fault tolerance techniques for distinct operators | adapt to operator properties |
| Optimized storage | [46] | computes an optimal number of checkpoints and levels | avoid exhaustive search |
| | [58] | utilization of solely relevant information | minimize stored information |

**Table 7**. A characterization of correlated checkpointing methods.

Balazinska et al. [132, 133] propose a fault tolerant approach to deal with node failures, network failures, and network partitions in the Borealis distributed stream processing system [131]. It is a replication-based method because distinct nodes run multiple copies of the same query network to ensure *availability* (i.e., deliver results within a specified time threshold). This method can tolerate *n* - 1 simultaneous node failures, if each node has *n* replicas. It also employs the upstream backup mechanism by buffering the tuples at the data sources. In this way, when failures occur, these tuples can be reprocessed to rebuild the operator's state. This method tries to ensure uninterrupted processing, despite failures, by continuing to process *tentative tuples* (i.e., tuples that belong to an input subset). These tentative tuples will be corrected later when failures heal, to produce a consistent result.

Chen et al. [20] checkpoint the entire system in order to ensure consistency. They employ *scalable coding strategies* to simultaneously handle multiple node or link failures. Unlike traditional fault tolerance schemes (i.e., performing a restart from a checkpoint), in this framework, applications are not aborted. Instead, they keep all of their surviving processes and adapt to the failures. Furthermore, they introduce several checkpoint encoding algorithms to improve scalability, such that "*the overhead to survive k failures in p processes does not increase as the number of processes p increases*."

Wang et al. [108] propose the *Meteor Shower* stream processing system, which utilizes *tokens* when checkpointing. As a first step, source operators initiate the flow of tokens throughout a streaming graph. Then, when an operator obtains these tokens, the system checkpoints the operator state. *Meteor Shower* is comprised of three techniques: (1) source preservation, to avoid the cost of handling redundant tuples in previous check-pointing mechanisms [49, 50, 57], (2) parallel and asynchronous checkpointing, to enable operators to keep running during the checkpointing process, and (3) application-aware checkpointing that can both adapt to changes to an operator's state size and checkpoint whenever the state size attains a minimum value. This method can handle both single and network failures.

Su et al. [100] develop a *passive and partially active* (PPA) scheme, to overcome weaknesses in fault tolerance methods (FTM). For example, *active* FTM require extra resources and *passive* FTM have a costly recovery process. The PPA scheme employs *passive checkpointing for all tasks* and partially-*active checkpointing for a selected number of tasks, since resources are limited*. Consequently, their scheme provides very fast recovery for a selected number of tasks that use active fault tolerance and tentative output for those tasks that exploit passive fault tolerance. Although the tentative output is less accurate than the exact output, its accuracy improves when more data are available. To generate the maximum quality of the tentative outputs, the PPA scheme employs a bottom-up dynamic programming algorithm to optimize the replication plan for correlated failures.

Upadhyaya et al. [105] propose using varying *fault-tolerance techniques for distinct operators* that correspond to a single query plan. Incidentally, such a strategy will require a cost-based optimization plan to achieve fault-tolerance. Thus, the authors introduce a fault-tolerance optimizer, called *FTOpt*, to automatically pair each operator with the most suitable technique in a query plan. *FTOpt* aims to reduce the execution time of the entire query despite failures. Their approach, like the *PPA* scheme, does not limit checkpointing to a single method. However, it is better than *PPA*, in terms of the quality of the result, since *FTOpt* produces exact output, as opposed to tentative output. Furthermore, this method can handle various kinds of failures (i.e., from process failures to network failures).

Hakkarinen et al. [46] propose an alternative approach, i.e., an *N-level diskless checkpointing* method that minimizes fault tolerance overhead, to cope with concurrent processor failures. In comparison to a one-level scheme, layering diskless checkpointing can enable failure tolerance up to a maximum of *N* processes and considerably reduces the runtime. In addition, the authors develop and verify an analytical cost model for diskless checkpointing. Lastly, their checkpointing scheme can also calculate the optimal number of checkpoints and levels, to avoid an exhaustive search.

Koldehofe et al. [58] propose a novel method that can survive multiple simultaneous node failures without using persistent checkpoints. They observe that "*at certain points in time, the execution of an event-processing operator solely depends on a distinct selection of events from the incoming streams, which are reproducible by predecessor operators*." This leads them to design a method that preserves the operator state in *savepoints,* instead of checkpoints. Consequently, the operator state solely requires the information necessary for the incoming streams and the relevant selection events. Their proposed savepoint recovery system can: (1) identify an empty operator state, (2) capture and replicate savepoints and ensure the reproducibility of corresponding events, and (3) tolerate multiple simultaneous operator failures.

### *4.3.3* Incremental Checkpointing

The approaches discussed in subsections 3.1 and 3.2 depend on the periodic checkpointing of state (PCoS) for failure recovery. However, Carbone et al. [18] discuss two key drawbacks. First, the PCoS often interrupts the overall computation, which slows down the data flow processing speed. Second, they greedily persist all tuples jointly with the operation states, thereby, resulting in larger than expected state sizes. Thus, to overcome these drawbacks, researchers propose methods based on incremental checkpointing [50, 96, 111], which only checkpoint changes to the state (not the entire state). By capturing the *delta* of the state (i.e., the latest changes in content, since the last checkpoint), these methods considerably reduce the checkpoint overhead and yield smaller state sizes. Table 9 contains a characterization of seven incremental checkpointing (IC) methods across systems. Next, we highlight the seven IC methods developed for use in the event of node failure.

Hwang et al. [50] propose a fine-grained checkpointing method that employs a divide-and-conquer strategy. In their scheme, the entire dataflow graph is divided into several subgraphs, each of which is then allocated to a different backup server. By employing a so-called *delta-checkpointing* technique, each server checkpoints a small fragment of its query graph. To guarantee state consistency, changes to state are incrementally checkpointed to backup servers. When failure occurs, query fragments are collectively recovered in parallel, thereby, achieving fast failure recovery and experiencing a small run-time overhead.

The *continuous eventual checkpointing* (CEC) method due to Sebepou et al. [96] guarantees fault tolerance by employing incremental state checkpoints continually, while minimizing interruptions to operator processors. To achieve this, operator state is split into parts and independently checkpointed, as needed. These *partial state checkpoints* are expressed as control tuples that contain the partial state of an operator. Unlike traditional schemes, in the CEC approach, checkpoints are updated incrementally and continuously. Consequently, the CEC method can efficiently handle continuous incremental state checkpoints and adjust check-point intervals to strike a balance between recovery time and running time.

| Common Characteristics | System | Main Mechanism | Objective |
|---|---|---|---|
| Divide-and-conquer strategy | [50] | checkpoint a small fragment of query graph | efficient checkpointing & failure recovery |
| | [96] | split operator state into control tuples | balance recovery and running time |
| | [111] | split states into slice units | fast recovery |
| Adapts to computing environments | [83] | utilize the similarity of access patterns | adapt to scarce memory |
| | [51] | multi-level checkpointing with delta compression | adapt to I/O & network bandwidth |
| N/A, only one system | [82] | compute the optimal number of incremental checkpoints | reduce checkpointing overhead |
| N/A, only one system | [18] | inject barriers into data | minimize space requirements |

**Table 9**. A characterization of incremental checkpointing methods.

Also employing a divide-and-conquer strategy, Wu et al. [111] propose *ChronoStream*, which splits states into a collection of fine-grained *slice units*. Comparable to the recently mentioned subgraph to backup server assignment scheme [50], units can be selectively distributed and checkpointed into specific nodes. Upon failure, ChronoStream transparently rebuilds the distributed slice units, and thus incurs small overhead. In comparison to other methods [50, 96], ChronoStream models application-level internal states differently.

To exploit the similarity of access patterns, among writes to memory in iterative applications, Nicolae et al. [83] propose the *Adaptive Incremental Checkpointing* (AI-Ckpt) approach for iterative computations under memory limitations. Under the assumption that "*first-time writes to memory generate the same kind of interference as they did in past iterations*," the AI-Ckpt method enables the prediction of future memory accesses for subsequent iterations. Consequently, this prediction leverages both current and historical access trends for flushing memory pages to stable storage in an optimal order. This asynchronous checkpointing approach is well suited for computing environments with limited memory resources. It can dynamically adapt to various applications, utilize access pattern history, and minimize the intervention of the checkpointing process running in the background.

Since the I/O and network bandwidth to distant storage heavily influences the checkpointing execution time (for large-scale systems), Jangjaimon et al. [51] propose the *adaptive incremental checkpointing* (AIC) method. This approach *reduces the state size*, to use bandwidth more efficiently, *lowers the overhead*, and *improves performance*. It employs multiple cores to perform adaptive multi-level checkpointing with delta compression, which can significantly minimize the incremental checkpoint file size. The authors also introduce a new Markov model to predict the performance of a multi-level concurrent checkpointing scheme. In comparison to checkpointing schemes employing fixed checkpoint intervals, the AIC method substantially reduces the expected running time (e.g., by 47%), when evaluated against six SPEC benchmarks.

Using a cost model, the incremental checkpointing (IC) method due to Naksinehaboon et al. [82] computes the *optimal number* of incremental checkpoints between two full checkpoints. Consequently, it reduces the checkpointing overhead, in comparison to full checkpoint (FC) models. Improving upon the IC approach, Paun et al. [90] extend it to include the *Weibull failure distribution* case. Their experiments show that the overhead of the IC method is significantly smaller than that of the FC method.

To minimize space requirements in dataflow execution engines, Carbone et al. [18] devise the *Asynchronous Barrier Snapshotting* (ABS) algorithm, which is suited for both acyclic and cyclic dataflows. On acyclic topologies, stage barriers, injected into data sources by a coordinator, can trigger the snapshot of current state. The algorithm solely materializes operator states in acyclic dataflows. On the other hand, on the cyclic execution graphs, ABS solely stores a minimal set of records on cyclic dataflows. Upon failure, the ABS algorithm reprocesses logged records to recover the system. Experiments show that ABS can achieve linear scalability and performs well with frequent state captures.

# 5. Integrative Optimization

Thus far, *state* has been shown to be effective in several isolated application scenarios (i.e., fault tolerance, load balancing, elasticity). However, state can also be used to simultaneously address multiple scenarios simultaneously (e.g., scalability, fault tolerance [36, 43, 80, 81, 111]). It is in this scenario that multi-objective or integrative optimization (IO) arises. For otherwise optimizing independently (per each scenario) would yield a suboptimal solution. Incidentally, IO spans numerous facets, as reflected in Table 10 (under Multipurpose).

Often, a single system alone cannot meet all processing requirements, such as *high-throughput batch processing*, *low-latency stream processing*, and *efficient iterative and incremental computations*. Therefore, multiple systems must be employed to achieve coverage. However, the use of a federation of platforms brings numerous problems, including inefficiency, complexity, and maintenance challenges. Hence, new systems are being developed with these multiple objectives in mind. Next, we discuss the varying IO methods prevalent across varying systems.

McSherry et al. [73] propose a new computational model, called *differential dataflow*, which supports both incremental and iterative computation. Extended from batch-oriented models (e.g., MapReduce, DryadLINQ), their model enables *arbitrarily nested* fixed-point iteration and simultaneously supports the efficient, incremental updates to inputs. Rather than using the entire temporal order, *changes to collections* are described in terms of the partial order. This allows the collections to evolve and eliminate the need to restart the computation to reflect changes.

Due to McSherry et al., *Naiad* [74, 79] is a distributed system for dataflow programs that is developed to satisfy all three of earlier referenced requirements in a single framework. Figure 11 presents Naiad at a high-conceptual level. Naiad supports both iterative and interactive queries on data streams and generates *up-to-date* and *consistent* results that can be incrementally updated, as new data arrive continuously. Furthermore, in [74, 79], McSherry et al., present a novel computational model, called *timely dataflow*, to boost the parallelism prevalent across various classes of algorithms (e.g., iterative, graph-based, tree-based).

| Common Characteristics | System | Main Mechanism | Multipurpose |
|---|---|---|---|
| Utilize state to address multiple problems simultaneously | [73] | differential dataflow | incremental and iterative-computation |
| | [74, 79] | timely dataflow | incremental, iterative-computation, high-throughput, and low-latency processing |
| | [36] | upstream backup | fault tolerance and scalability |
| | [111] | transactional migration protocol & thread-to-slice mapping | fault tolerance, scalability, and elasticity |
| | [43] | partition functions | load balancing and operator migration |
| | [80] | reuse checkpoints for load balancing | fault tolerance, state migration, and load balancing |
| | [81] | mixed-integer linear programs | load balancing and scalability |

**Table 10**. A characterization of integrative optimization methods.

To describe the logical points during execution, Naiad employs *timestamps* to enhance dataflow computation. Timestamps are essential in supporting an efficient and lightweight coordination mechanism. This is due to three features, namely, structured loops for feedback, stateful dataflow vertices for records processing (without using global coordination) and notifying vertices when all tuples have been received by the system for a specific input or iteration round. While the first two features support low-latency iterative and incremental computation, the third feature ensures the result is consistent.

Fernandez et al. [36] develop a unified approach based on stateful dataflow graphs (SDG) for dynamic scalability and failure recovery, to parallelize stateful operators (when workloads fluctuate) and achieve fast recovery times (with low overhead). In their approach, they use the upstream backup to periodically checkpoint stateful operators. Their system detects bottlenecks in operators and enables them to scale by automatically allocating new machines. Consequently, repartitioning the state, accordingly. In the event of a failure, the checkpointed state will need to be rebuilt on a new machine and tuples will need to be reprocessed to recover the failed operators. To achieve these goals, the proposed system: (i) uses a well-defined interface to allow for the easy access to operator state, (ii) reflects information about the exact set of processed tuples by an operator in its state, and (iii) preserves operator semantics using a key attribute to partition the state.

Wu and Tan's *ChronoStream* [111] concurrently offers fault tolerance, scalability, and elasticity. Their low-latency stream processing system provides transparent workload reconfiguration in a unified model, by separating application-level parallel computation (i.e., computation states) from OS-level execution concurrency. As a result, ChronoStream achieves transparent elasticity, fault tolerance, and high availability without having to sacrifice performance. This is due to the reduction in the overhead triggered by state synchronization. The slice-reconstruction approach in ChronoStream is akin to the state-migration approach in SEEP [36]. Furthermore, both Wu and Tan's ChronoStream and SDG [37] support dynamic reconfiguration at runtime. However, state repartitioning incurs high state migration costs in both SEEP and SDG.

We revisit the method of Gedik et al. [43] to address both load balancing and operator migration. Recall that their solution employs a partitioning function, to achieve improved load balance (auto-fission) and low migration costs. The structure of the partitioning function is a hybrid involving an explicit map and a consistent hash. Consequently, this compact hash function can balance the workload uniformly and adapt accordingly, even under high skew. Furthermore, they construct algorithms and metrics to build and assess the partitioning functions, to determine whether these can achieve good balance and efficient migration. More precisely, they define *load imbalance* to be the proportion of the difference between the maximum and minimum loads to the maximum permissible load difference. Data items in the partially constructed partitioning function have their migration costs normalized based on the ideal migration cost. The *utility function* combines the relative imbalance metric and the migration cost metric, to assign the items to parallel channels.

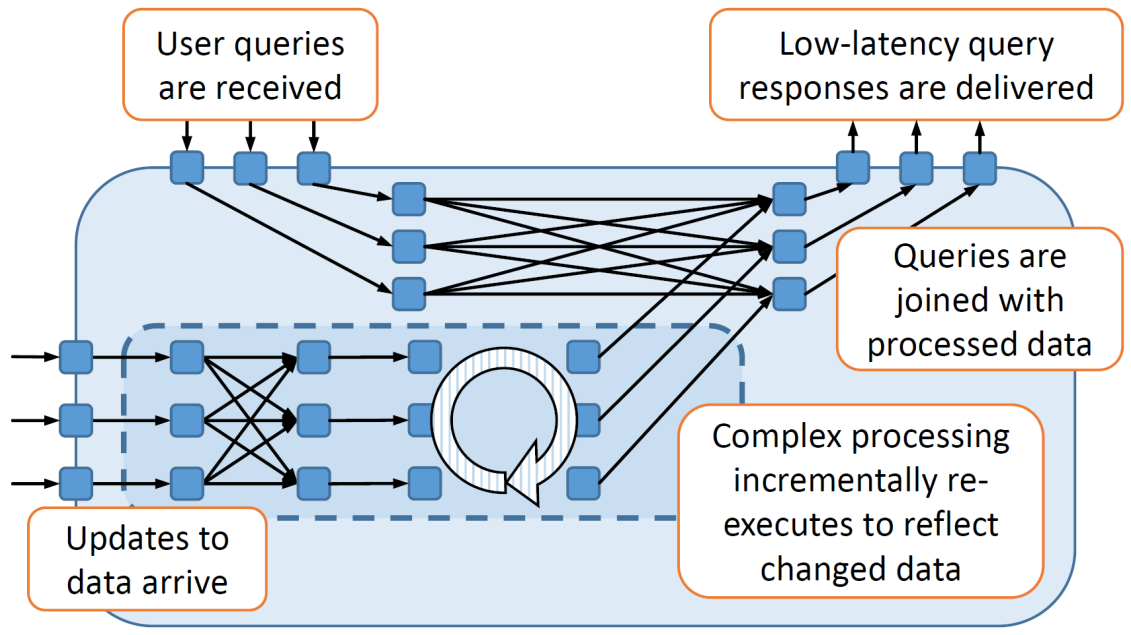

**Figure 11**. A *Naiad* application that answers real-time queries on continuously updated data [74]. Iterative, incremental processing is represented by the dashed rounded rectangle.

Madsen et al. [80] re-use the checkpoints meant for failure recovery, to efficiently improve the dynamic migration of the state, like Fernandez et al. [36]. As a first step, they formally define a checkpoint allocation problem with some constraints. Then they propose a practical (i.e., efficient) algorithm to reuse the checkpoints for effective load balancing. If the workload is increasingly skewed at key groups, then the system must transfer many checkpoints, for groups of keys in $A$, where $A$ is a set of key groups, to nodes with lighter loads in advance, to quickly react to fluctuations. To increase the chance of this availability, the checkpoints of the key groups in $A$ must be allocated to the nodes with key groups that are "*as negatively as possible correlated with the key groups of A.*" Due to the relationship between fault tolerance and migration, checkpointing can be viewed as *proactive load-balancing*, i.e., utilizing checkpoints for state migration to help balance the load.

Lastly, Madsen et al. [81] model *load balancing*, *operator instance placement*, and *horizontal scaling*, jointly, to enable low-latency processing, optimize resource usage, and minimize communication costs. They integrate horizontal scaling and load balancing using *mixed-integer linear programs* (MILP) to arrive at a feasible solution. This model is suitable when the placement of operator instances does not considerably affect communication costs. By using the MILP approach and linear program solvers, they improve the load balance over existing heuristic approaches. Yet, using the so-called *Autonomic Load Balancing with Integrated Collocation* (ALBIC) solution enables them to further achieve gains over the MILP based approach. Using ALBIC, they can: (i) generate an improved operator instance collocation, (ii) balance the load, and (iii) lower the overhead. This holds because ALBIC gradually improves the placement at runtime, while still satisfying load balance constraints.

# 6. Implementations of State and Limitations

In this section, we survey the implementations of state in five popular open-source big data processing frameworks, i.e., *Storm*, *Heron*, *Samza*, *Spark*, and *Flink*. Table 11 summarizes some of the characteristics corresponding to each of these frameworks. Next, we examine the varying system implementations and highlight some of their limitations.

| Systems | State Management | Fault Tolerance | Guarantees |
|---|---|---|---|
| **Storm** | not native | tuples acknowledge | *at least once* |
| **Storm + Trident** | specific operators | tuples acknowledge | *exactly once* |
| **Heron** | stateful topologies | tuples acknowledge | *at least once* |
| **Samza** | stateful operators | log of updates | *at least once* |
| **Spark** | state DStream | RDD lineage | *exactly once* |
| **Flink** | stateful operators | State checkpoint | *exactly once* |

**Table 11**. Implementation of state across systems.

Storm solely supports stateless processing and implements state management at the application level, to enable fault-tolerance and scalability in stateful applications. It is not equipped with any native mechanism to manage state. To overcome this limitation, an abstraction layer called *Trident* that extends Storm has been proposed. It is a micro-batch system that adds state management and guarantees exactly-once semantics using its own API designed for fault tolerance. It not only inherits Storm's acknowledgement mechanism, it can prevent data loss and ensure that each tuple is processed only once.

Currently, there are two state management alternatives supported in Storm. The first alternative keeps information about both the order of the most recent batch and the current state, however, it may block execution. The second alternative overcomes the previously stated shortcoming, however, it incurs more overhead, by also maintaining the last known-state. To ensure *correct semantics*, it is vital to maintain the order of state updates. Storm provides *at-least-once* guarantees by re-emitting tuples from a spout (i.e., a data source) in the event of failure. It uses an upstream backup technique and tuple acknowledgements to reprocess tuples in the event of failure. In contrast, Storm + Trident provides *exactly-once* guarantees by writing topologies with required semantics. To achieve these semantics, Trident uses three primitives: (1) tuples are processed in small batches, (2) each batch is assigned a unique id called the *transaction id*, unless the batch is being replayed, in which case the batch is given the same id, and (3) state updates are ordered among batches, i.e., state updates for batch $i+1$ must wait until the state updates for batch $i$ are complete. However, Trident is ill-suited for big states, for otherwise, it would incur severe delays.

Storm has a number of limitations. First, it is hard to debug. Second, it requires a special hardware allocation, which limits its scalability. Finally, it requires the manual isolation of machines when managing provisioning. Thus, Kulkarni et al. [130] proposed Heron to overcome these limitations. Heron uses stateful topologies comprised of spouts and processing elements (i.e., bolts). In these topologies, every component, both spouts and bolts store their state when processing tuples. Like Storm, Heron uses tuple acknowledgements for fault tolerance (i.e., each tuple in the system is acknowledged, once it is fully processed by downstream components). Heron can deliver *at most once* guarantees (without acknowledgement) or *at least once* guarantees (when employing acknowledgement).

Samza can manage large states (e.g., GBs in each partition) by preserving state in local storage and using Kafka to duplicate state changes. Kafka stores the log of state updates and can easily restore state. By default, Samza uses a key-value store to support stateful operators. However, alternative storage systems are also available, if richer querying capabilities are required. Like Storm, Samza offers *at-least-once* guarantees in the event of failure by re-emitting messages.

Spark implements state management using the concept of a DStream (i.e., a discretized stream), which updates operations via transformations. Distributed immutable collections or RDDs (resilient distributed datasets) are key concepts of Spark. Fault tolerance in Spark is achieved using lineage [116], to avoid checkpointing overhead. *State* in Spark streaming plays the role of another micro-batching stream. For this reason, during micro-batch processing, Spark uses an old *state* to generate another micro-batch result and a new *state*.

Specifically, the transform function separates state from the output, enabling programmers to call RDD functions on micro-batches. Then, they can use functions, such as RDD.join(), to combine the state with incoming tuples. Spark achieves *exactly-once* semantics in one of two ways, i.e., either idempotent writes or transactional writes. In idempotent writes (i.e., multiple writes that produce the same data), messages are stored in a database, according to a unique key without duplication. In transactional writes, messages are written to storage within a single transaction. Due to this atomic operation, transaction rollbacks eliminate duplicated messages.

Flink employs a single-pass algorithm that superimposes global snapshotting to normal execution [18], to support exactly-once semantics. This approach is akin to the Chandy-Lamport algorithm, which uses markers. However, unlike the Chandy-Lamport algorithm, which assumes a strongly connected distributed system, this Flink-specific algorithm also applies to weakly connected execution graphs. To checkpoint state, Flink offers a wide range of configurable state backends, with various levels of complexity and persistence.

Currently, Flink keeps state in memory (i.e., holds state internally as objects on the Java heap), backs up state in a file system (e.g., HDFS), or persists state in RocksDB. Flink also introduces the concept of *queryable state* [125], which enables real-time queries to directly access event-time windows, thereby avoiding the overhead associated with writing to key/value stores. Consequently, with these enhancements to *state*, Flink can also support many other operations, such as software patches, testing, and system upgrades. Like Spark, Flink uses idempotent and transactional writes to support *exactly-once* semantics [125].

Although the implementations of *state* differ in these frameworks, in terms of their *representation* and *storage solutions*, they all lack support for adaptive checkpointing. As of the time of the writing of this survey, these frameworks solely support periodic checkpointing (e.g., hourly checkpointing). Some researchers [94] prove that aperiodic checkpointing can improve performance over periodic check-pointing. Thus, one appealing research direction is to extend these frameworks to support adaptive checkpointing (i.e., determine when to optimally checkpoint adaptively as opposed to checkpointing periodically). We can calculate these optimal moments using the checkpointing (and recovery) costs at the time checkpoints happen. These costs, in turn, depend on the probability that failures occur. Consequently, this cost-based adaptive checkpointing model must integrate the anticipation of failure probability as an important parameter. Additionally, devising an efficient representation of state (e.g., *approximate*, *compressed*, *incrementally-updateable*) that enables iterative algorithms to run more efficiently is yet another opportunity for further research.

# 7. Open Discussions

We conclude this survey by motivating new research directions in state management. This includes novel approaches to: (1) integrate state management into big data frameworks, (2) enable state management for iterative algorithms, (3) use state to support hybrid systems, and (4) evaluate state management methods.

## 7.1 Integrating State Management into Big Data Frameworks

Current big data frameworks can be further extended to incorporate existing techniques for state management at varying abstraction levels, ranging from *low-level* (e.g., operator primitives, calculus algebra) to *high-level* (e.g., language level or platform level). Next, we discuss each level in greater detail.

At the *lowest* level, primitive operators can be further extended, beyond what was discussed in subsection 3.1.5. By incorporating leading state management solutions into the current frameworks, managing state will be far easier to do and lead to greater efficiencies.

At the *calculus* level, researchers [17, 34, 47] focus on incremental state computation using algebra. Cai et al. [17] introduce a new mathematical theory (i.e., the theory of *changes* and *derivatives*) for incremental computation. Hammer et al. [47] use *first-class names* as the essential linguistic characteristic for efficient incremental computation. Fegaras [34] uses *monoid homomorphisms* as the underlying mechanism to propose an algebra for distributed computing. Consequently, at the algebraic level, we can extend the incremental change of state to support additional functions, beyond those discussed in this survey.

At the *high-language* level, some researchers [7, 99] devise novel declarative languages for big data processing. Silva et al. [99] propose a language to allow users to easily define and parameterize checkpointing policies. In this framework, *language annotations* are used to apply fault tolerance policies in streaming applications. Further, this approach combines language primitives with code generation to facilitate checkpointing per user specification. Beyond fault tolerance, language annotation extensions (LAE) can specify parts of an application that should be actively vs. passively (e.g., PPA scheme [100]) fault-tolerant. Additionally, LAE may be used to declare which operators should be made public (e.g., for users) vs. private (e.g., for internal operator use only). Furthermore, Alexandrov et al. [7] propose the *Emma* language, which *deeply embeds* APIs into a host language (e.g., Scala) for complex data analysis. Emma can be further extended to integrate state management methods at the language level, thereby enabling *declarative state management*.

At the *high platform* level, *Rheem* [3, 4] introduce multi-layer (i.e., *platform*, *core*, and *application*) data processing and a storage abstraction to support both *platform independence* and *inter-operability* across platforms. They envision that a data processing abstraction based on user-defined functions can achieve two purposes. First, users can solely focus on the logic of their data analytics tasks. Second, applications can be independent from data processing platforms. Rheem decomposes a complex analytic into smaller subtasks to leverage diverse processing platforms. This division allows a single task to run over multiple platforms to boost performance. Moreover, we can further extend Rheem to build a state management system that eases deployment on various platforms, achieves independence and interoperability among platforms, and improves performance.

In this subsection, many perspectives were presented. Some researchers have already begun to incorporate high-level support for declarative big data analysis. However, determining how to combine the strengths of each of these individual systems, in order to support state management declaratively remains a challenging research problem.

## 7.2 Enabling State Management for Iterative Algorithm Based Applications

Many machine learning algorithms, such as PageRank, K-Means, and its variants [110] require iterative steps to converge to the final solution. Due to big state sizes, some iterative algorithms use *approximate state* with small sizes [1, 2, 42, 52] or *approximate algorithms* with fewer iterative steps [16, 65, 77, 88, 113, 119, 120] to boost performance. Usually, these approximate algorithms sacrifice accuracy for performance. However, some researchers [39, 40, 41, 115] develop solutions that ensure both precision and performance. These approaches investigate mechanisms to represent state in an approximate form, approaches for optimizing approximate algorithms, and the development of exact iterative algorithms. Ultimately, these solutions focus on increasing performance with minimal impact result quality.

Seamlessly and efficiently incorporating approximate state representations into the exact algorithms is another challenging problem. Once this has been achieved, we can then compare the approximate and exact algorithms, in terms of precision and performance to determine how state approximation can help boost latency and/or throughput.

For emerging application scenarios, such as the Internet of Things (IoT), continuous data streams must be processed with very short delays. Determining how to use state efficiently in these applications to satisfy the abovementioned requirement is a challenging problem. For example, Hochreiner et al. [129] propose a platform for stream processing in the IoT, where they use synchronized state across all computing nodes. They also provide a toolkit for developers to manage shared state.

## 7.3 Using State Management for Hybrid Systems

While batch data provides comprehensive and historical views of data, real-time streaming data provides fresh and up-to-date information. Some researchers [14, 75, 76] propose hybrid systems to process these two types of data on a single platform. These hybrid systems handle both historical information and the most recent data.

The Lambda architecture [71] tries to process both batch and streaming data by providing a software stack including: (1) a batch layer (e.g., implemented in Hadoop) to process batch data, (2) a speed layer (e.g., implemented in Storm) to process streaming data, and (3) a serving layer to index batch views and enable them to be queried in low-latency. This mixture of multiple systems is hard to configure, manage, and maintain due to their diversity and heterogeneity. Moreover, many data analysis tasks generally involve both layers, thereby limiting optimization opportunities. Thus, we cannot process data as efficiently as a single unified system.

To partially overcome this weakness in the Lambda architecture, Jay Kreps proposes the Kappa architecture [138], which removes the batch layer and only uses a single stream processing engine. However, Kappa is not a perfect replacement for Lambda, especially in situations, where batch and streaming algorithms have differing outputs (e.g., for machine learning). Other researchers [14, 75, 76] propose hybrid systems that integrate multiple data types (e.g., real-time with batch or streaming with OLAP). Boykin et al. [14] propose *Summingbird* to combine online and batch MapReduce computations into a single framework. To fuse stream and transaction processing into a single system, Meehan et al. [75] built *S-Store*. Initially, starting with a completely transactional OLTP database system, then integrating additional streaming functionality. This enables S-Store to simultaneously and seamlessly support OLTP and streaming applications. Meehan et al.'s [76] *BigDAWG*, tightly integrates stream and batch processing, to enable seamless and high-performance querying capability over both new and historical data. The effectiveness of BigDAWG in practical applications is discussed in Elmore et al. [30].

Systems, such as S-Store and Summingbird do not directly focus on combining batch and streaming data in a single system. Consequently, future research can encapsulate the entire functionality of a Lambda architecture into a single system to take advantages of both batch and streaming worlds. Then devising novel state checkpointing methods is an essential requirement for stateful hybrid applications. Moreover, proposing new ways to manage state in incremental computations for both batch and streaming data in a single framework is an intriguing research problem. Batch and streaming data have specific characteristics. Thus, additional research will need to be conducted to develop novel methods for efficient state management that meets both batch and streaming data requirements.

## 7.4 Evaluating State Management Methods

Evaluating *state management* methods is of paramount importance. However, deciding *which evaluation criteria or standards* to use is still an open problem. There are numerous state management methods, but no universal benchmark (with associated datasets, metrics, and workloads) that are widely accepted. As a starting point, we propose the following four dimensions to consider.

- *Efficiency*: State management methods should have low latency and high performance, particularly, when considering state updates, state migration, and state purging. For example, this could be attained by efficient algorithms that exploit compression or approximate/ incomplete storage. Performance metrics may include *state size*, *accuracy*, and *precision* when using approximate state during computations, and traditional performance metrics, such as *latency* and *throughput*.

- *Ease of Use/Management*: APIs that use and access state must be simple and easy to use. They should cover most application scenarios and provide richer functions and encapsulations. This will help to reduce the human latency cost in deploying and using big data frameworks in the future. User studies could serve as an evaluation method to assess the expressiveness and effectiveness of state management APIs for a variety of problem domains.

- *Functionality*: Evaluating the functionality and adequateness of state management for a particular application is another important dimension. For example, state can efficiently support iterative algorithms in many different domains, such as artificial intelligence and machine learning. For efficiency, it may have to support multiple consistency guarantees and allow users to choose which consistency level to use during a given iteration. This type of functionality may not be supported by certain state management APIs. Comparing and relating different functionalities may guide a user to select the appropriate state management systems and methods.

- *Seamless Integration*: New methods should easily integrate into existing, ongoing, and future frameworks for big data processing. The integration must be effective, i.e., not requiring too much effort to modify existing, underlying platforms and not imposing any impedance mismatch.

## 8. Conclusion

In this survey, we have analyzed and surveyed *state management* research in big data processing systems from two perspectives. Broadly speaking, we have taken a closer look at the varying concepts of state management, we have discussed a variety of methods that one could use to operate on state, including storing, updating, purging, migrating, and exposing.

We have presented varying approaches for *incrementally maintaining* state, to prevent full state maintenance, which is expensive. We have highlighted how state could be *shared* among operations, to more efficiently utilize resources and reuse computational results. We have demonstrated how state ensures *elasticity and load balance* in parallel or distributed big data processing systems by flexibly splitting work-loads among computing nodes. Moreover, we have looked at state management performance considerations, including overhead and complexity related issues.

In addition, we have reviewed the varying applications of state. For example, how state is used to enable *stateful computation*, to support complex operations that combine state and input. We have illustrated how state is used to accelerate *iterative computation* via the reuse of state values. We also have showcased how state is used to enable *fault tolerance* and facilitate failure recovery.

We have highlighted how state can be utilized to enable *multiple optimizations goals* in a single system simultaneously (e.g., incremental and iterative computation, fault tolerance and elasticity). Furthermore, we have compared state implementation among five popular frameworks. Unfortunately, none of these frameworks has addressed all of the abovementioned issues. Thus, reducing the complexity, lowering processing latency, and enabling fast recovery remain active research areas. Today, researchers [94] are actively working on solving these problems and in this endeavor *adaptive checkpointing* seems to hold promise.

Finally, in Appendix A, we list state management research contributions by dimension. We hope this survey will pave the way for subsequent state management research (e.g., state integration and approximation, state usage in hybrid systems, evaluation metrics) for big data processing systems.

## Acknowledgments

This work was funded by the H2020 STREAMLINE project under grant agreement No 688191 and by the German Federal Ministry for Education and Research (BMBF) funded Berlin Big Data Center (BBDC), under funding mark 01IS14013A.

## Appendix A. State Management *Concepts & Applications of State* by Approach / Paper / System

| *Approach* / Research Paper / *System* | Stateful Computation | Fault Tolerance | Iterative Processing | Elasticity & Load Balance | Integrative Optimization | State Sharing | Incremental Maintenance | Operations | Overhead Complexity |
|---|---|---|---|---|---|---|---|---|---|
| 1. *Flink* [134] | ✓ | ✓ | ✓ | ✓ | | | | | |
| 2. *Spark* [137] | ✓ | ✓ | | | | | | | |
| 3. *Heron* [135] | | ✓ | | | | | | | |
| 4. *Samza* [136] | ✓ | ✓ | | | | | | | |
| 5. Logothetis et al. [67] | ✓ | | | | | | | ✓ | |
| 6. Logothetis et al. [66] | ✓ | | | | | | | | |
| 7. Gedik et al. [43] | ✓ | | | ✓ | ✓ | | | | |
| 8. Matteis et al. [72] | ✓ | | | | | | | | |
| 9. *CEC* [96] | | ✓ | | | | | | | |
| 10. Hwang et al. [50] | | ✓ | | | | | | | |
| 11. *Scalable Coding Strategies* [20] | | ✓ | | | | | | | |
| 12. *Meteor Shower* [108] | | ✓ | | | | | | | |
| 13. Koldehofe et al. [58], Hakkarinen et al. [46] | | ✓ | | | | | | | |
| 14. *PPA* [100] | | ✓ | | | | | | | |
| 15. *FTOpt* [105] | | ✓ | | | | | | | |
| 16. *AI-Ckpt* [83] | | ✓ | | | | | | ✓ | |
| 17. Naksinehaboon et al. [82] | | ✓ | | | | | | | ✓ |
| 18. Paun et al. [90] | | ✓ | | | | | | | |
| 19. *AIC* [51] | | ✓ | | | | | | | |
| 20. *ABS* [18] | | ✓ | | | | | | | |
| 21. Ewen et al. [31, 32] | | | ✓ | | | | | | |
| 22. Schelter et al. [95], Dudoladov et al. [27] | | ✓ | ✓ | | | | | | |
| 23. *head & tail checkpoint* [112] | | ✓ | ✓ | | | | | | |
| 24. *MRQL Streaming* [33] | | | ✓ | | | | ✓ | ✓ | |
| 25. *Flux* [98] | | | | ✓ | | | | | |
| 26. *ChronoStream* [111] | | ✓ | | ✓ | ✓ | | | ✓ | |
| 27. *differential dataflow* [73] | | | | | ✓ | | | | |
| 28. *Naiad* [74] | | | | | ✓ | | | | |
| 29. Fernandez et al. [36] | | ✓ | | | ✓ | | | ✓ | ✓ |
| 30. Madsen et al. [80, 81] | | | | | ✓ | | | | |
| 31. Brito et al. [15] | | | | | | ✓ | | | |
| 32. Gordon et al. [44], Arasu et al. [9] | | | | | | ✓ | | | |
| 33. Sermulins et al. [97] | | | | | | ✓ | | | |
| 34. Kuntschke et al. [59] | | | | | | ✓ | | | |
| 35. *CAPSULE* [68] | | | | | | ✓ | | | |
| 36. *S-Store* [75], [101] | | | | | | ✓ | | | |
| 37. *ring of databases* [54] | | | | | | | ✓ | | |
| 38. *viewlet transforms* [5] | | | | | | | ✓ | | |
| 39. *DBToaster* [55] | | | | | | | ✓ | | |
| 40. *LINVIEW* [84] | | | | | | | ✓ | | |
| 41. Nikolic et al. [85], Koch et al. [56] | | | | | | | ✓ | | |
| ~~42. Liu et al. [64]~~ | | | | | | | ~~✓~~ | | |
| 43. Zhang et al. [118] | | | | | | | | ✓ | |
| 44. Liu et al. [63] | | | | | | | | ✓ | |
| 45. *SGuard* [57] | | ✓ | | | | | | ✓ | |
| 46. *CALC* [92] | | | | | | | | ✓ | |
| 47. *Photon* [8] | | | | | | | | ✓ | |
| 48. *SDG* [37] | | | | | | | | ✓ | |
| *49.* *GraphLab* [69] | | | | | | | | ✓ | |
| 50. *PJoin* [25] | | | | | | | | ✓ | |
| 51. *Punctuation semantics* [104] | | | | | | | | ✓ | |
| 52. Li et al. [61], Zhu et al. [121] | | | | | | | | ✓ | |
| 53. Ding et al. [26] | | | | | | | | ✓ | ✓ |
| 54. *StreamCloud* [45] | | | | | | | | ✓ | |
| 55. *SBON* [91] | | | | | | | | ✓ | |
| 56. *MigCEP* [86] | | | | | | | | ✓ | |
| 57. *MLCBF* [35] | | | | | | | | ✓ | |
| 58. Sayed et al. [94] | | | | | | | | | ✓ |
| 59. Robert et al. [93] | | | | | | | | | ✓ |
| 60. Bouguerra et al. [13] | | | | | | | | | ✓ |